\pdfoutput=1
\documentclass[sigconf, nonacm]{acmart}

\usepackage{amsmath,amsfonts}
\usepackage[ruled,vlined]{algorithm2e}
\usepackage{graphicx}
\usepackage{textcomp}
\usepackage{xcolor}
\usepackage{tabularx}
\usepackage{booktabs}
\usepackage{makecell}
\usepackage{ragged2e}
\usepackage{pgfplots, pgfplotstable, siunitx}
\usepackage{multirow} 
\usepackage{threeparttable} 
\usepackage{url}

\usepackage{enumitem,mdframed}
\newlist{compactitem}{itemize}{1}
\setlist[compactitem]{nosep,leftmargin=*,label=\textbullet,itemsep=1pt,topsep=0pt}

\definecolor{cardbg}{RGB}{249,250,252} 
\definecolor{cardbd}{RGB}{214,219,226} 
\newmdenv[
  linewidth=0.5pt,
  linecolor=cardbd,
  backgroundcolor=cardbg,
  roundcorner=5pt,
  innerleftmargin=8pt,
  innerrightmargin=8pt,
  innertopmargin=6pt,
  innerbottommargin=6pt,
]{MergePipecard}

\usepackage{fancyhdr}

\usepackage{xcolor}
\usepackage{listings}
\usepackage{caption}

\usepackage{subcaption}
\usepackage{wrapfig}

\graphicspath{{./fig/}}

\newcommand\vldbdoi{XX.XX/XXX.XX}

\newcommand\vldbavailabilityurl{URL_TO_YOUR_ARTIFACTS}

\begin{document}
\title{MergePipe: A Budget-Aware Parameter Management System for Scalable LLM Merging}

\author{Yuanyi Wang}
\affiliation{%
  \institution{Hong Kong Polytechnic University}
}
\email{yuan-yi.wang@connect.polyu.hk}

\author{Yanggan Gu}
\affiliation{%
  \institution{Hong Kong Polytechnic University}
}
\email{yanggangu@outlook.com}

\author{Zihao Wang}
\affiliation{%
  \institution{InfiX.ai, Hong Kong}
}
\email{wangzihao2020@lzu.edu.cn}

\author{Kunxi Li}
\affiliation{%
  \institution{Zhejiang University}
}
\email{kunxili@zju.edu.cn}

\author{Yifan Yang}
\affiliation{%
  \institution{Hong Kong Polytechnic University}
}
\email{yi-fan.yang@connect.polyu.hk}

\author{Zhaoyi Yan}
\affiliation{%
  \institution{InfiX.ai, Hong Kong}
}
\email{yanzhaoyi@outlook.com}

\author{Congkai Xie}
\affiliation{%
  \institution{InfiX.ai, Hong Kong}
}
\email{xiecongkai@infix-ai.com}

\author{Jianmin Wu}
\affiliation{%
  \institution{Hong Kong Polytechnic University}
}
\email{jianmin.wu@polyu.edu.hk}

\author{Hongxia Yang}
\affiliation{%
  \institution{Hong Kong Polytechnic University}
}
\email{hongxia.yang1@gmail.com}


\begin{abstract}
Large language model (LLM) merging has become a key technique in modern LLM development pipelines, enabling the integration of multiple task- or domain-specific expert models without retraining.
However, as the number of experts grows, existing merging implementations treat model parameters as unstructured files and execute merges in a stateless, one-shot manner, leading to excessive disk I/O, redundant parameter scans, and poor scalability.

In this paper, we present \textbf{MergePipe}, a parameter management system for scalable LLM merging.
MergePipe is the first system that treats LLM merging as a data management and execution problem, and introduces a catalog-driven abstraction over model parameters, merge plans, and execution lineage.
At its core, MergePipe employs a cost-aware planner that explicitly models expert parameter I/O and enforces user-specified I/O budgets, followed by a streaming execution engine that materializes merged models under transactional guarantees.
Our key insight is that while base model reads and output writes are unavoidable, expert parameter reads dominate merge cost and constitute the primary optimization target.
By making expert access budget-aware throughout planning and execution, MergePipe mitigates the $O(K)$ I/O growth of naive pipelines and achieves predictable scaling behavior.
Experiments show that MergePipe reduces total I/O by up to an order of magnitude and delivers up to $11\times$ end-to-end speedups (up to 90\% wall-time reduction) over state-of-the-art LLM merging pipelines.

\end{abstract}

\maketitle




\section{Introduction}
\label{sec:intro}
Large language model (LLM) development has increasingly shifted from training a single monolithic, dense model to composing models from multiple expert checkpoints or branches \citep{naveed2025comprehensive}.
Existing LLMs are often maintained as families of checkpoints, consisting of a general-purpose base model and numerous domain-specialized experts (e.g., for mathematics \citep{shao2024deepseekmath,yang2024qwen2}, code \citep{hui2024qwen2,guo2024deepseek}, law \citep{yue2024lawllm,cui2023chatlaw}, or biomedicine \citep{xu2025lingshu}), trained at the scale of billions to trillions of tokens \citep{yang2025qwen3,agarwal2025gpt}.
Rather than restarting training from scratch, contemporary development pipelines increasingly rely on \textbf{model merging} to combine such expert checkpoints, enabling multi-skill integration, domain adaptation \citep{wang2025model,sung2023empirical}, and rapid iteration with modest additional compute \citep{bidermanlora,lu2024twin}.
Unlike ensembling, which retains multiple models and aggregates predictions online at the cost of increased latency and memory \citep{ganaie2022ensemble}, model merging operates directly on parameters to produce a single deployable model.
As a result, LLM merging has been widely adopted in recent large-scale models, including DeepSeek-V3 \citep{liu2024deepseek} and Qwen3 \citep{yang2025qwen3}, where merging is used to extend training, integrate specialized capabilities, or consolidate intermediate models during development.

\begin{figure}[t]
\centering
\includegraphics[width=0.99\linewidth]{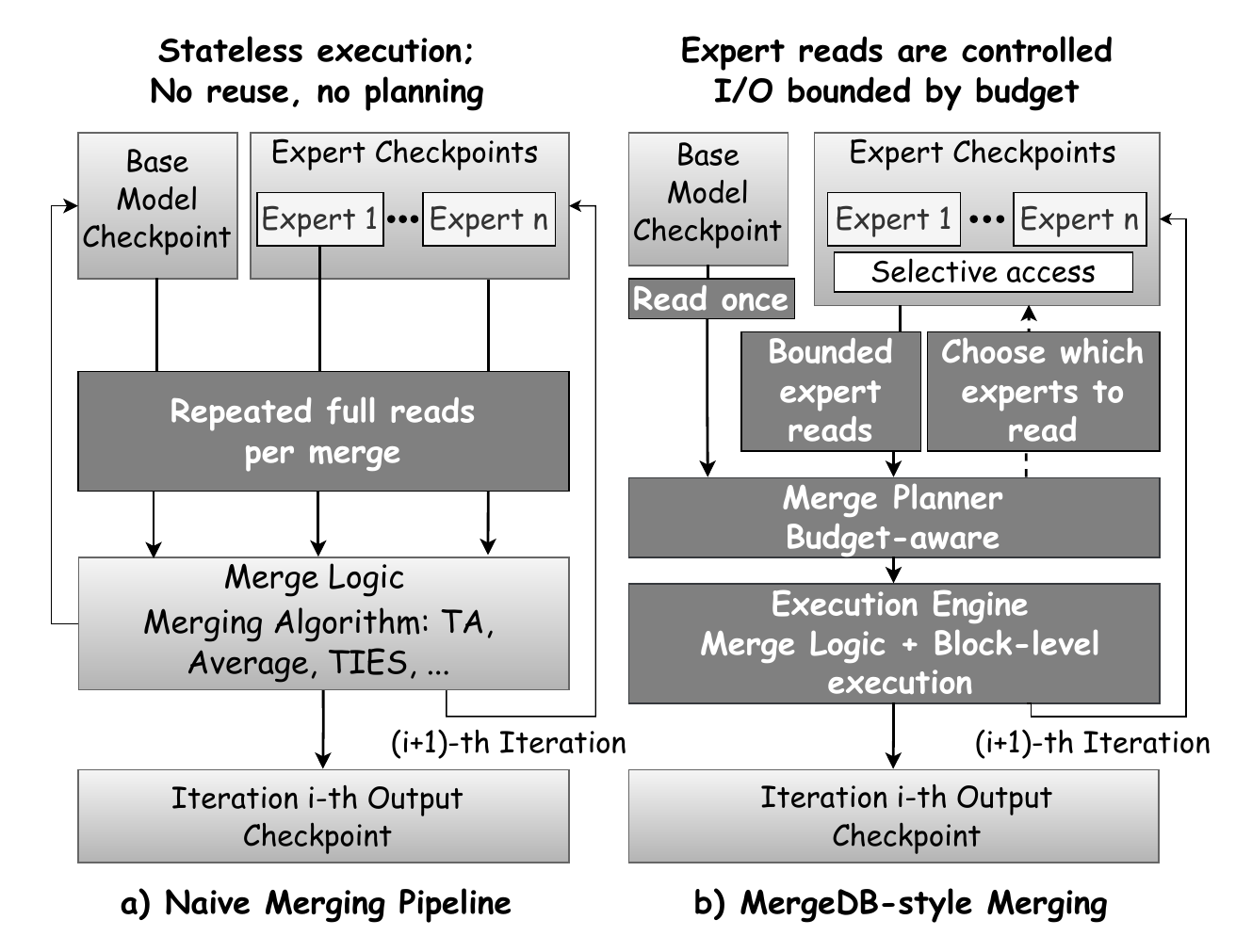}
\caption{Naive LLM merging vs.\ MergePipe. Naive pipelines merge in a stateless, one-shot manner and repeatedly scans expert checkpoints ($O(K)$ expert I/O), while MergePipe introduces planning, reuse, and budget-bounded expert reads.}
\label{fig:overview}
\end{figure}

\begin{figure}[t]
\centering
\includegraphics[width=0.98\linewidth]{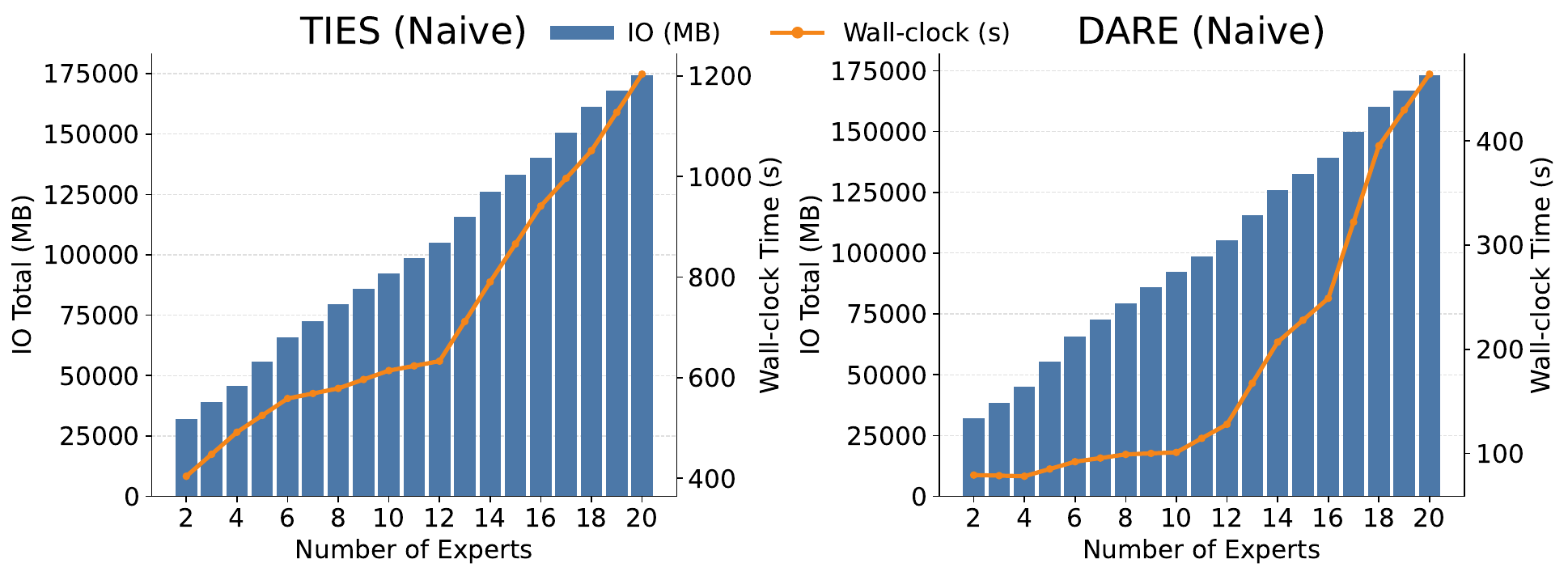}
\caption{\textbf{Naive merging scales poorly with experts.}
For 3B models, both TIES and DARE show near-linear growth in total I/O (bars) as $K$ increases, and wall time rises accordingly.}
\label{fig:naive_merge}
\end{figure}

Despite its growing importance, existing LLM merging pipelines are largely implemented in a system-agnostic manner.
Most current implementations treat model parameters as unstructured flat files and execute merging as a stateless, one-shot process \citep{yang2024model, lu2024merge}.
Each merge invocation independently loads expert parameters, applies the merging logic, and materializes the output, without reuse across merges, explicit lineage tracking, or any notion of cost-aware planning.
As a result, expert parameters are repeatedly scanned in full, even when merging the same base model with an increasing set of experts.
In modern LLM pipelines, where checkpoints can span tens to hundreds of billions of parameters, this design causes merge I/O cost to scale linearly with the number of experts, quickly rendering large-scale merging expensive and impractical.
Importantly, these inefficiencies stem not from the merging algorithms themselves, but from the absence of a system-level abstraction that treats parameters, merge plans, and execution state as first-class entities.
Fig.~\ref{fig:overview} and Fig.~\ref{fig:naive_merge} illustrate this contrast at the pipeline level, highlighting how naive merging repeatedly scans expert parameters across merge iterations.
Specifically, this calls for database-style primitives over parameters: a persistent catalog over parameter blocks, declarative merge plans with explicit cost semantics, and transactional snapshotting with lineage for reproducibility.

To understand why such execution models scale poorly, we examine the underlying cost structure of LLM merging.
The end-to-end cost of a merge can be decomposed into three components: (i) reading the base model parameters, (ii) materializing the merged output checkpoint, and (iii) reading expert parameters.
The first two components are intrinsic to the semantics of model merging, as any merge must access the base model and produce a new checkpoint.
In contrast, expert parameter access exhibits fundamentally different behavior.
Under naive execution, expert checkpoints are accessed independently for each merge invocation, causing expert I/O cost to grow linearly with the number of experts~$K$.
This $O(K)$ scaling quickly dominates the overall merge cost and accounts for the majority of data movement in large-scale merging workloads.
This decomposition exposes a crucial asymmetry for system design: \textit{base reads and output writes are semantic necessities, whereas the dominant expert-read cost is an execution artifact of stateless pipelines and can be controlled via planning and reuse.}
In other words, the cost of expert is the only term whose growth with $K$ is not mandated by merge semantics, making it the primary optimization target.
Consequently, controlling expert parameter access, rather than optimizing merge computation, becomes the primary strategy for scalable LLM merging.

However, simply recognizing expert parameter access as the dominant cost is not sufficient to achieve scalable merging in practice.
Expert access patterns vary substantially across merging algorithms and workloads.
Some merging algorithms uniformly operate on all parameters, like model soup \citep{wortsman2022model} and TA (Task Arithmetic) \citep{ilharcoediting}, while others selectively access expert parameters based on conflict detection or similarity thresholds, like TIES \citep{yadav2023ties} and DARE \citep{yu2024language}.
As a result, the potential benefit of reducing expert I/O differs widely across merging strategies, and cannot be captured by fixed heuristics or ad-hoc caching mechanisms.
Therefore, effective optimization requires a principled way to model, predict, and reason about expert access cost under different merging plans.

Moreover, introducing reuse, consistency, and lineage tracking into the merging process inevitably incurs system-level overheads.
While such overheads may outweigh optimization benefits in small-scale merging scenarios, they become amortized as the number or size of expert checkpoints grows.
This trade-off implies that no single execution strategy is optimal across all regimes.
Instead, LLM merging calls for a system that can explicitly expose and control these trade-offs, allowing users to balance expert access and execution overheads according to workload characteristics.

To address these challenges, we propose \textbf{MergePipe}, a parameter management system designed for scalable and reproducible LLM merging.
Rather than introducing a new merging algorithm, MergePipe redefines model merging as a data management problem and provides a database-backed abstraction over model parameters, merge plans, and execution state.
In MergePipe, merge plans are treated as \emph{first-class objects} that explicitly capture which parameters are accessed, how they are combined, and under what resource constraints.
This design decouples parameter storage, merge planning, and execution, enabling principled reuse across merges and explicit control over expert access.
Specifically, MergePipe introduces a cost-aware planning framework that models expert parameter I/O as an explicit, budgeted resource.
Given a user-specified I/O budget, the planner derives execution plans that selectively schedule expert access while preserving merge semantics.
These plans are enforced by an execution engine that provides transactional writes and fine-grained lineage tracking, ensuring correctness, reproducibility, and incremental reuse across merge iterations.

In summary, this work reveals three key insights:
\begin{compactitem}
\item \textbf{I1: Expert reads dominate scalability rather than merge computation.}
Naive pipelines repeatedly scan expert checkpoints across invocations, causing expert I/O to grow nearly linearly with the number of experts $K$ (Fig.~\ref{fig:overview}, Fig.~\ref{fig:naive_merge}).

\item \textbf{I2: Expert I/O is the primary controllable cost knob in LLM merging.}
Base model reads and output writes are intrinsic to merge pipelines, while scalable merging hinges on reducing expert reads via reuse and budget-aware access (Fig.~\ref{fig:fig7_vldb_all}).

\item \textbf{I3: Budget-aware planning enables predictable and operator-agnostic control.}
By modeling expert parameter access as an explicit resource, MergePipe enforces user-specified I/O budgets while generalizing across diverse merge operators (Table~\ref{tab:merge-generality}).
\end{compactitem}

Our contributions are:
(1) MergePipe, the first \emph{parameter management system} for iterative LLM merging, which provides a catalog-driven data model over parameter blocks, merge plans, and execution lineage;
(2) a budget-aware planning framework that binds the cost model to executable plans, enforcing an explicit expert-read budget to eliminate the $O(K)$ expert-scan behavior of naive pipelines;
(3) a execution engine with transactional snapshotting and fine-grained lineage, enabling reproducible and auditable materialization of merged checkpoints under the planned budget.

\begin{figure*}[t]
  \centering
  \includegraphics[width=\textwidth]{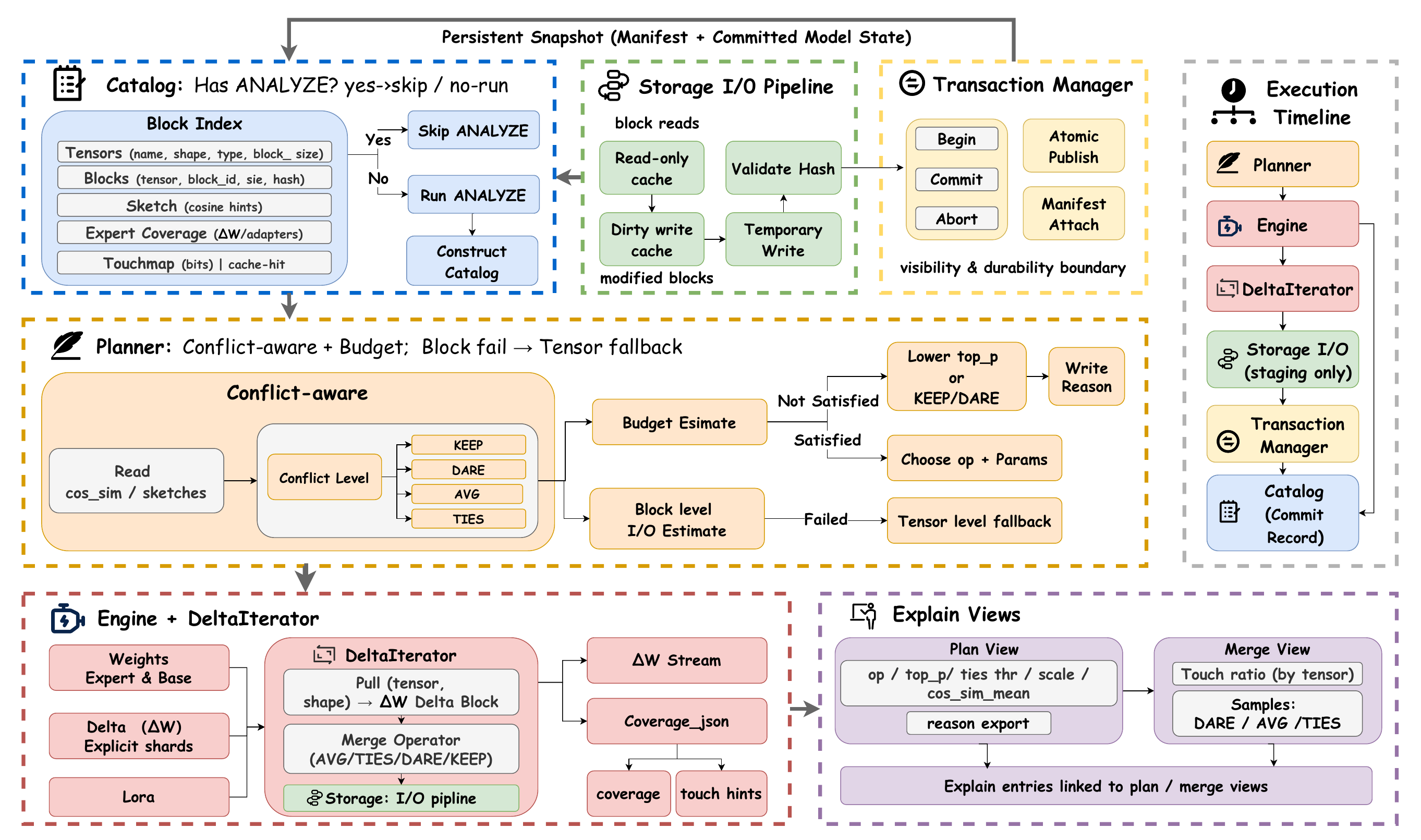}
  \caption{\textbf{MergePipe system overview.}
MergePipe decouples parameter storage, budget-aware planning, and streaming execution for scalable LLM merging.
A persistent catalog supports block-level indexing/reuse; a cost- and conflict-aware planner selects expert blocks under an explicit I/O budget; and an execution engine enforces the plan via \texttt{DeltaIterator} with atomic publish and lineage/explainability records.}
  \label{fig:MergePipe-overview}
\end{figure*}

\section{System Overview}
\label{sec:overview}

To address the scalability and reproducibility challenges in large-scale LLM merging (Section~\ref{sec:intro}), we present \textbf{MergePipe}, a parameter management system that treats merging as a data management workload. Instead of optimizing specific merge algorithms, MergePipe targets the underlying execution model by decoupling parameter storage, merge planning, and execution.
Figure~\ref{fig:MergePipe-overview} overviews the system: (i) a persistent catalog and storage I/O pipeline, (ii) a cost- and conflict-aware planner, (iii) a streaming execution engine, and (iv) transactional and explainability components for lineage. Together, they enable budgeted expert access, reuse across iterations, and auditable merges.

\subsection{Design Goals and Scope}
\label{sec:design-goals}

MergePipe targets \emph{iterative} merging where a slowly evolving base model is repeatedly merged with a growing set of experts under constrained I/O. Guided by the cost structure in Section~\ref{sec:cost-model}, we adopt:

\noindent
\textbf{G1: Bounded expert I/O.}
Expert reads dominate merge cost and grow as $O(K)$ in naive pipelines; MergePipe
treats expert access as a budgeted resource.

\noindent
\textbf{G2: Explicit planning before execution.}
MergePipe separates planning from execution. Plans specify accessed parameters and resource constraints, enabling cost estimation, reuse, and budget-aware fallback.

\noindent
\textbf{G3: Persistent execution.}
MergePipe persistently records decisions, access, and lineage to support auditability, reuse, and recomputation.

\noindent
\textbf{Scope:}
MergePipe manages and executes merging without introducing new merge algorithms. It is operator-agnostic such as TIES, DARE and supports full weights, parameter deltas, and adapter-based experts. We focus on disk-dominated regimes where parameter I/O is the bottleneck, not training-time in-memory fusion.

\subsection{Data Model and Catalog}
\label{sec:catalog}

MergePipe is built around a persistent catalog that manages model parameters and merge-related metadata as first-class data. The catalog provides (i) block-granular views for costing and reuse, and (ii) immutable snapshots with manifests for reproducible execution.

\noindent
\textbf{Block:}
A model checkpoint $M$ consists of named tensors. Each tensor $T$ is partitioned
by a deterministic function $\textsc{Partition}(T; s)$ into fixed-size blocks
$\{b\}$, where $s$ is the block size. A \emph{block id} is a stable identifier
$\langle \textit{model\_id}, \textit{tensor\_id}, \textit{block\_idx}\rangle$ that
uniquely locates a physical block in storage.

\noindent
\textbf{Snapshot and Manifest:}
A committed merge produces a \emph{snapshot} identified by $sid$ and a
\emph{manifest} $\mathsf{man}(sid)$.
$\mathsf{man}(sid)$ is an immutable mapping from each tensor-block key $\langle \textit{tensor\_id}, \textit{block\_idx}\rangle$ to a storage object reference (plus execution metadata such as inputs, operator, and plan digest). Snapshots are the unit of reuse, inspection, and incremental recomputation.

\noindent
\textbf{Invariant (Immutability and Atomic Visibility).}
Committed snapshots are immutable: once $\mathsf{man}(sid)$ is published, its
block mapping and metadata never change. Each merge execution either atomically
publishes a new snapshot $sid'$ with a complete manifest, or publishes nothing
(i.e., no partial visibility).

\noindent
\textbf{Core catalog records.}
To support block-level cost estimation, reuse detection, and auditable lineage,
MergePipe stores a small set of structured records. Table~\ref{tab:catalog-schema}
summarizes the core relations (logical schema; the physical backend can be a
KV-store or embedded DB).

\begin{table}[t]
\centering
\small
\caption{Core catalog records used by MergePipe for block-level planning, reuse,
and reproducible snapshots.}
\setlength{\tabcolsep}{4pt}
\resizebox{\linewidth}{!}{
\begin{tabular}{l p{0.72\linewidth}}
\toprule
\textbf{Record} & \textbf{Key fields (PK underlined)} \\
\midrule
\textsc{BlockMeta} &
\underline{model\_id}, \underline{tensor\_id}, \underline{block\_idx};
bytes, shape, dtype, hash, sketch, layout \\
\textsc{TouchMap} &
\underline{sid}, \underline{tensor\_id}; touched\_blocks (bitmap / ranges) \\
\textsc{Coverage} &
\underline{sid}, \underline{tensor\_id}, \underline{block\_idx};
expert\_set\_digest (or list) \\
\textsc{Plan} &
\underline{plan\_id}; base\_id, expert\_ids, op, budget\_B,
selected\_blocks\_digest, $\widehat{C}_{\text{expert}}$ \\
\textsc{Manifest} &
\underline{sid}; plan\_id, base\_id, expert\_ids, op, budget\_B,
(realized) $C_{\text{expert}}$, output\_root, created\_at \\
\bottomrule
\end{tabular}}
\label{tab:catalog-schema}
\end{table}

\noindent
\textbf{How the catalog enables planning and reuse.}
\textsc{BlockMeta} provides the minimal statistics needed for block-level cost
accounting and lightweight conflict signals. \textsc{Plan} records the
selected expert blocks under a budget and can be reused across iterative merges.
\textsc{TouchMap} and \textsc{Coverage} summarize what was actually materialized and which experts contributed, enabling explainability and incremental recomputation. \textsc{Manifest} binds a plan to an immutable snapshot boundary, making merges reproducible and auditable.

\subsection{Merge Planning and Execution Workflow}
\label{sec:workflow}

MergePipe follows a staged workflow over \emph{immutable snapshots}
(Section~\ref{sec:catalog}). \texttt{ANALYZE} and planning are metadata-only, while execution atomically publishes a new snapshot
(Fig.~\ref{fig:MergePipe-overview}).

\noindent
\textbf{Analysis.}
MergePipe checks the catalog for existing analysis results; on misses, it
computes and persists block-level metadata like shapes, hashes, sketches, and
coverage for reuse.

\noindent
\textbf{Planning.}
Given a merge request and expert I/O budget, the planner estimates expert access costs from catalog metadata and produces a budget-feasible plan at block granularity when possible (with tensor-level fallback).

\noindent
\textbf{Execution and commit.}
The engine enforces the plan by streaming blocks through \texttt{DeltaIterator},
incrementally applying merge operators and materializing only selected blocks.
Writes are staged and atomically published via the transaction manager, while touch maps and coverage summaries are recorded for explainability and reuse.
By separating analysis, planning, and execution, MergePipe achieves predictable resource usage, incremental reuse, and reproducible merging without constraining the merge operators.

\section{Cost Model for Model Merging}
\label{sec:cost-model}

This section introduces a cost model for large-scale LLM merging.
The goal of the model is not to predict exact execution time, but to capture
the dominant cost factors that determine scalability under constrained I/O
resources.
The model provides a principled abstraction that underlies MergePipe's planner
and execution design.

\subsection{Overview and Assumptions}

We consider the problem of merging a base model with a set of expert models.
Let
\[
\mathcal{M} = \{ M_0, M_1, \ldots, M_K \}
\]
denote the models involved in a merge, where $M_0$ is the base model and
$M_1, \ldots, M_K$ are expert models.
Each model consists of a collection of tensors:
\[
M_i = \{ T_{i,1}, T_{i,2}, \ldots, T_{i,n_i} \}.
\]
Each tensor is stored persistently on disk; we denote its physical size by
$\text{size}(T)$.

We focus on disk I/O cost as the dominant bottleneck in large-scale merging.
This assumption is consistent with empirical observations that merging
workloads involve reading and writing tens to hundreds of gigabytes of
parameters, while computation remains comparatively lightweight.
A key distinction in MergePipe is between costs that are intrinsic to the
semantics of model merging and costs that can be controlled through planning
decisions.
Only the latter are subject to optimization.

\subsection{Cost Decomposition}

We decompose the total cost of a merge operation as:
\[
C_{\text{merge}} =
C_{\text{base}} +
C_{\text{expert}} +
C_{\text{out}} +
C_{\text{meta}}.
\]
The individual components are defined as follows.

\noindent
\textbf{Base Model Read Cost.}
The base model must be fully read to construct the merged model:
\[
C_{\text{base}} = \sum_{T \in M_0} \text{size}(T).
\]
This cost is invariant across merge strategies and independent of the number
of experts.

\noindent
\textbf{Output Materialization Cost.}
The merged model must be fully materialized and written to storage:
\[
C_{\text{out}} = \sum_{T \in M_{\text{merged}}} \text{size}(T).
\]
The merged model preserves the tensor structure of the base model.

\noindent
\textbf{Metadata Cost.}
Metadata operations include lineage recording, hashing, and transactional
commit:
\[
C_{\text{meta}} = C_{\text{lineage}} + C_{\text{hash}} + C_{\text{commit}}.
\]
These are bounded and weakly dependent on the merge strategy.

\noindent
\textbf{Expert Read Cost.}
The dominant and controllable component is the expert read cost:
\[
C_{\text{expert}} =
\sum_{i=1}^{K} \sum_{T \in \mathcal{S}_i} \text{size}(T),
\]
where $\mathcal{S}_i \subseteq M_i$ denotes the subset of tensors accessed from expert $M_i$.
Under naive merging pipelines, all expert tensors are read:
\[
C_{\text{expert}}^{\text{naive}} =
\sum_{i=1}^{K} \sum_{T \in M_i} \text{size}(T),
\]
which scales linearly with the number of experts $K$.
In contrast, MergePipe aims to reduce $C_{\text{expert}}$ by selectively accessing
the expert paramaters.

\subsection{Block-Level Cost Estimation}

MergePipe refines expert cost estimation by operating at block granularity.
Each tensor $T$ is partitioned into a set of fixed-size blocks
$T = \{ b_1, \ldots, b_m \}$.
The expert read cost can be expressed as:
\[
C_{\text{expert}} =
\sum_{i=1}^{K} \sum_{b \in \mathcal{B}_i} \text{size}(b),
\]
where $\mathcal{B}_i$ denotes the selected blocks from expert $M_i$.

Tensor-level estimation is a special case of this formulation in which all
blocks of a tensor are selected.
Block-level estimation enables finer-grained control over expert access and
serves as the basis for MergePipe's planning and fallback mechanisms.

\subsection{Budget-Constrained Planning Objective}

MergePipe introduces an explicit I/O budget constraint on the expert reads:
\[
C_{\text{expert}} \leq B,
\]
where $B$ is a user-specified budget (like in MB).

The budget applies exclusively to expert reads.
Base model reads, output materialization, and metadata operations are treated
as fixed overheads and are not subject to optimization.
Under this constraint, the planner seeks to select expert blocks
$\{ \mathcal{B}_i \}$ that respect the budget while following merge-specific
heuristics:
\[
\min \; C_{\text{expert}}
\quad \text{s.t.} \quad
C_{\text{expert}} \leq B.
\]

The optimization problem is not solved exactly.
Instead, the cost model provides a structured abstraction that guides
budget-aware planning heuristics, described in
Section~\ref{sec:planner}.
By separating fixed and controllable costs, the model enables predictable
scaling behavior as the number of experts increases: while naive pipelines
exhibit $O(K)$ expert I/O growth, MergePipe bounds I/O independently of $K$.

\section{Merge Planning}
\label{sec:planner}

This section describes the design of MergePipe’s planner, which generates budget-aware merge plans based on the cost model in Section~\ref{sec:cost-model}.
The planner’s objective is to control expert parameter access while preserving the semantics of the specified merge operator.
Rather than optimizing merge computation, the planner explicitly reasons about expert I/O cost and enforces user-specified budgets.

\subsection{Planner Interface and Scope}

The planner takes as input:
(i) a merge request specifying a base model, a set of expert models, and a merge
operator like AVG, TIES, DARE;
(ii) a user-specified expert I/O budget $B$;
and (iii) catalog metadata produced by prior \texttt{ANALYZE} phases
(Section~\ref{sec:catalog}).

The planner outputs a \emph{merge plan} that declaratively specifies:
(a) which expert parameters (or blocks) are accessed,
(b) which merge operator and parameters are applied, and
(c) the expected expert read cost.
The plan is a first-class object that can be inspected, reused, and executed
independently by the execution engine.

\noindent
\textbf{Planner scope.}
The planner does \emph{not} modify merge formulas or introduce new merge
algorithms.
Instead, it determines \emph{which parameters are accessed} under constrained
resources.
This separation ensures that MergePipe remains agnostic to merge operators while
providing predictable system-level behavior.
We formalize a merge plan as an explicit selection-and-execution contract.

\begin{definition}[Merge Plan]
A merge plan is a tuple
\[
\pi = \big(\mathrm{op}, \theta, \{\mathcal{B}_i\}_{i=1}^{K}, \mathrm{order}\big),
\]
where $\mathrm{op}$ is the merge operator, $\theta$ denotes operator parameters, $\mathcal{B}_i$ is
the set of selected parameter blocks to be accessed from expert $M_i$, and
$\mathrm{order}$ is a deterministic tensor or block traversal order used by the execution engine.
\end{definition}

\begin{definition}[Feasibility]
\label{definiation2}
Let $\widehat{C}_{\text{expert}}(\pi) = \sum_{i=1}^{K}\sum_{b\in \mathcal{B}_i}\text{size}(b)$
be the planned expert read cost under block-level estimation.
A plan $\pi$ is \emph{budget-feasible} under expert I/O budget $B$ iff
\[
\widehat{C}_{\text{expert}}(\pi) \le B.
\]
\end{definition}

\subsection{Cost Binding to the Model}

The planner instantiates the cost model in Section~\ref{sec:cost-model} by binding each candidate plan $\pi$ to an estimated expert read cost $\widehat{C}_{\text{expert}}(\pi)$.
Only $\widehat{C}_{\text{expert}}(\pi)$ participates in feasibility checks, while $C_{\text{base}}$, $C_{\text{out}}$, and $C_{\text{meta}}$ are treated as
fixed overheads.
A plan is accepted only if it satisfies the budget-feasibility condition in Definition \ref{definiation2}.

\subsection{Conflict-Aware Signals}

To prioritize expert parameters under budget constraints, the planner leverages
conflict-aware signals derived from catalog metadata.
These include lightweight sketches (e.g., cosine similarity hints) and coverage
statistics computed during \texttt{ANALYZE}.
Intuitively, these signals estimate how much an expert parameter diverges from the base model or overlaps with other experts.
The planner uses these signals solely for \emph{ranking} and \emph{selection}.
They do not alter merge operator semantics.
For example, under AVG merging, selected parameters are still averaged; under TIES or DARE, the same thresholding or masking rules apply.
Conflict-aware signals only influence \emph{which parameters are accessed} when
resources are limited.

\subsection{Budget-Aware Plan Generation}

MergePipe generates plans using a greedy procedure that incrementally selects
expert blocks under the explicit budget $B$.
Algorithm~\ref{alg:plangen} formalizes this process: the planner (i) assigns a
priority score $s(i,b)$ to each candidate block using catalog signals (e.g.,
sketches and coverage), (ii) iterates candidates in descending score, and (iii)
adds a block to the plan iff the incremental cost preserves feasibility,
i.e., $\text{cost}+\text{size}(b)\le B$.
When the next candidate would violate the budget, the planner either skips the
block or (when supported by the operator) applies a bounded adjustment to
operator parameters (e.g., decreasing a threshold $\theta$ for TIES/DARE);
all such decisions are recorded in the plan metadata for reproducibility.
The resulting plan $\pi$ is budget-feasible by construction and maximizes the
use of high-priority expert information within $B$.

\begin{algorithm}[t]
\DontPrintSemicolon
\SetKwInOut{Input}{Input}
\SetKwInOut{Output}{Output}
\SetAlgoLined
\Input{Experts $\{M_i\}_{i=1}^K$; catalog $\mathcal{C}$; operator $\mathrm{op}$; budget $B$}
\Output{A budget-feasible plan $\pi$}

Compute a priority score $s(i,b)$ for each candidate expert block $b$ using catalog sketches/coverage\;
Sort candidates in descending $s(i,b)$\;
Initialize $\mathcal{B}_i \leftarrow \emptyset$ for all $i$, and cost $\leftarrow 0$\;
\ForEach{candidate $(i,b)$ in sorted order}{
  \If{$\text{cost}+\text{size}(b) \le B$}{
    $\mathcal{B}_i \leftarrow \mathcal{B}_i \cup \{b\}$;\ $\text{cost} \leftarrow \text{cost}+\text{size}(b)$\;
  }{
    \textit{optionally adjust $\theta$ for $\mathrm{op}$ and continue, otherwise skip}\;
  }
}
Return $\pi = (\mathrm{op},\theta,\{\mathcal{B}_i\},\mathrm{order})$\;
\caption{\textsc{PlanGen}: greedy budget-aware plan.}
\label{alg:plangen}
\end{algorithm}


\subsection{Fallback, Robustness, and Complexity}

When block-level metadata are missing or unreliable, the planner falls back to a
coarser tensor-level selection, i.e., selecting whole tensors (equivalently, all
blocks within selected tensors) and estimating $\widehat{C}_{\text{expert}}(\pi)$
using the tensor-level formulation in Section~\ref{sec:cost-model}.
Fallback events and their causes are recorded in the plan metadata for reproducibility.
Let $N_b$ denote the number of parameter blocks considered during planning.
The planner runs in $O(N_b \log N_b)$ time due to priority ordering and budget
checks.
In practice, planning overhead is negligible compared to execution I/O and is
amortized across iterative merge workflows via plan reuse.

\begin{algorithm}[t]
\DontPrintSemicolon
\SetAlgoLined
\SetKwInOut{Input}{Input}
\SetKwInOut{Output}{Output}
\SetSideCommentRight
\SetNoFillComment

\Input{Merge plan $\pi$; base model $M_0$; experts $\{M_i\}_{i=1}^K$; storage $\mathcal{S}$; catalog $\mathcal{C}$; transaction manager $\mathcal{T}$}
\Output{Committed snapshot id $sid$ and manifest $\mathsf{man}$}

\BlankLine
\textbf{Transaction and staging.}\;
$\mathcal{T}.\textsc{Begin}()$\;
$w \leftarrow \mathcal{S}.\textsc{OpenStagingWriter}()$\;
$\textit{touch} \leftarrow \emptyset$;\ $\textit{coverage} \leftarrow \emptyset$\;

\BlankLine
\textbf{(1) Stream selected blocks under plan $\pi$.}\;
\ForEach{tensor $t \in \pi.\textsc{TensorOrder}()$}{
  $D \leftarrow \textsc{InitDeltaIterator}(t,\pi,M_0,\{M_i\})$\;
  \ForEach{block $b \in \pi.\textsc{BlocksToMaterialize}(t)$}{
    $\{\Delta_i\} \leftarrow D.\textsc{Pull}(b)$ \tcp*[r]{expert blocks iff selected}
    $x_0 \leftarrow \mathcal{S}.\textsc{ReadBaseBlock}(M_0,t,b)$\;
    $x \leftarrow \textsc{ApplyOperator}\!\left(x_0,\{\Delta_i\},\pi.\textsc{Op}(t,b)\right)$\;
    $w.\textsc{WriteBlock}(t,b,x)$\;
    $\textit{touch}[t] \leftarrow \textit{touch}[t] \cup \{b\}$\;
    $\textit{coverage}[t,b] \leftarrow D.\textsc{UsedExperts}()$\;
  }
}
\BlankLine
\textbf{(2) Validate and atomically publish snapshot.}\;
$\mathcal{S}.\textsc{ValidateHashes}(w)$\;
$\mathsf{man} \leftarrow \mathcal{C}.\textsc{BuildManifest}(\pi,\textit{touch},\textit{coverage})$\;
$sid \leftarrow \mathcal{T}.\textsc{AtomicPublish}(w,\mathsf{man})$\;
$\mathcal{C}.\textsc{CommitRecord}(sid,\mathsf{man})$\;
$\mathcal{T}.\textsc{Commit}()$\;

\Return $(sid,\mathsf{man})$\;

\caption{\textsc{ExecuteMerge}: Budget-enforced streaming execution with atomic commit.}
\label{alg:executemerge}
\end{algorithm}

\section{Execution Engine}
\label{sec:engine}

MergePipe’s execution engine enforces a planner-produced merge plan $\pi$ and
materializes a merged \emph{snapshot} under an explicit expert I/O budget.
Algorithm~\ref{alg:executemerge} defines the execution semantics.

\subsection{Execution Semantics and Budget Soundness}

\paragraph{Definition (Realized expert reads).}
Let $\mathrm{Sel}_\pi(t,b)\subseteq\{1,\ldots,K\}$ be the experts selected by $\pi$
for output block $(t,b)$. Algorithm~\ref{alg:executemerge} realizes the expert
read set
\begin{align}
\mathcal{R}_{\text{expert}}(\pi)=\{(i,t,b)\mid i\in \mathrm{Sel}_\pi(t,b)\}, 
\\ C_{\text{expert}}^{\text{run}}(\pi)=\sum_{(i,t,b)\in \mathcal{R}_{\text{expert}}(\pi)} \mathrm{size}(b).
\end{align}
The output snapshot is a \emph{complete checkpoint}: for every base block $(t,b)$,
the engine reads $x_0$ and writes an output block $x$ (Algorithm~\ref{alg:executemerge},
Step~(1)); only expert blocks are selectively read via $D.\textsc{Pull}(b)$.

\paragraph{Property (Budget soundness).}
In Step~(1), the engine calls $D.\textsc{Pull}(b)$ \emph{iff} $(i,t,b)\in\mathcal{R}_{\text{expert}}(\pi)$,
thus the realized expert I/O is bounded by the plan and the user budget:
\[
C_{\text{expert}}^{\text{run}}(\pi)\ \le\ \widehat{C}_{\text{expert}}(\pi)\ \le\ B,
\]
up to accounting granularity used by the storage layer, like block-sized rounding in buffered I/O.
This property follows directly because $\pi$ explicitly enumerates all expert
blocks the engine is allowed to access, and the engine does not perform any
additional expert reads beyond those selected by $\pi$.

\subsection{DeltaIterator and Operator Application}

To support heterogeneous experts (full weights, deltas, adapters), the engine
uses a unified \texttt{DeltaIterator}. For tensor $t$, \textsc{InitDeltaIterator}
constructs $D$ such that $D.\textsc{Pull}(b)$ returns exactly the selected expert
contributions $\{\Delta_i\}$ for block $b$ (Algorithm~\ref{alg:executemerge},
Step~(1)). The engine computes each output block by
\[
x \leftarrow \textsc{ApplyOperator}\!\left(x_0,\{\Delta_i\},\pi.\textsc{Op}(t,b)\right),
\]
without modifying operator semantics; $\pi$ only controls \emph{which} expert
information is accessed.

\subsection{Atomic Publish and Manifested Lineage}

All writes are staged and hash-validated before publication
(Algorithm~\ref{alg:executemerge}, Step~(2)). The engine then builds a manifest
$\mathsf{man}\leftarrow \mathcal{C}.\textsc{BuildManifest}(\pi,\textit{touch},\textit{coverage})$
and atomically publishes the new snapshot id $sid$ via
$\mathcal{T}.\textsc{AtomicPublish}(w,\mathsf{man})$.

\paragraph{Property (Atomic visibility).}
Algorithm~\ref{alg:executemerge} guarantees that a merge either commits a single
new snapshot $(sid,\mathsf{man})$ or leaves no externally visible partial state.

\section{Experiments}
\label{sec:experiments}

The goal of this section is to empirically validate the key insights identified in Sections~\ref{sec:intro} and~\ref{sec:cost-model}, and to demonstrate that MergePipe delivers predictable and scalable performance by explicitly controlling expert parameter access.
Specifically, our experiments are organized around the following insights:
\begin{compactitem}
  \item \textbf{I1. Expert I/O dominates scalability:}
  Naive merging pipelines repeatedly scan expert checkpoints, causing expert I/O and end-to-end time to grow nearly linearly with the number of experts.

  \item \textbf{I2. Expert I/O is the primary controllable cost knob:}
  While base model reads and output writes are unavoidable, selectively reducing expert reads is the key to scalable merging without modifying merge semantics.

  \item \textbf{I3. Budget-aware planning enables predictable control:}
  Explicitly modeling expert I/O as a budgeted resource allows our MergePipe to
  generalize across different merge operators and model families.

  \item \textbf{I4. System overheads are small and amortized:}
  The additional costs introduced by planning, metadata management, and transactional execution are negligible relative to end-to-end merging and become amortized at realistic scales.

  \item \textbf{I5. Block granularity admits a robust operating regime:} MergePipe’s performance is stable across a practical range of block sizes and execution settings, indicating that its benefits do not rely on fragile tuning.
\end{compactitem}

All experiments are conducted without modifying the merging algorithms themselves,
and directly reflect the cost decomposition and budget-constrained planning objective introduced in Section~\ref{sec:cost-model}, with a particular focus on
controlling the expert read cost $C_{\text{expert}}$.
\footnote{Our project is provided in \url{https://github.com/wyy-code/mergepipe}.}

\begin{table}[t]
\renewcommand\arraystretch{1}
\setlength{\tabcolsep}{4pt}
\centering
\large
\caption{Evaluated models configuration in experiments.
All models use official HuggingFace checkpoints.
BF16 size is estimated assuming 2 bytes per parameter and reflects the approximate scale of parameter I/O during merging.
}
\resizebox{\linewidth}{!}{
\begin{tabular}{lccccc}
    \toprule
    \textbf{Model} & \textbf{Params} & \textbf{Layers} & \textbf{Hidden Dim} & \textbf{BF16 Size} & \textbf{Experts Number} \\
    \midrule
    Qwen3-0.6B      & 0.6B & 24 & 1024 & $\sim$1.2\,GB  & 20 \\
    Qwen3-1.7B      & 1.7B & 28 & 2048 & $\sim$3.4\,GB  & 20 \\
    Qwen3-8B        & 8.0B & 32 & 4096 & $\sim$16.0\,GB & 20 \\
    Llama-3.2-3B    & 3.0B & 26 & 3072 & $\sim$6.0\,GB & 25 \\
    Llama-3.1-8B    & 8.0B & 32 & 4096 & $\sim$16.0\,GB & 20 \\
    \bottomrule
  \end{tabular}}
\label{tab:model-specs}
\end{table}

\begin{figure*}[t]
  \centering
  \includegraphics[width=0.95\textwidth]{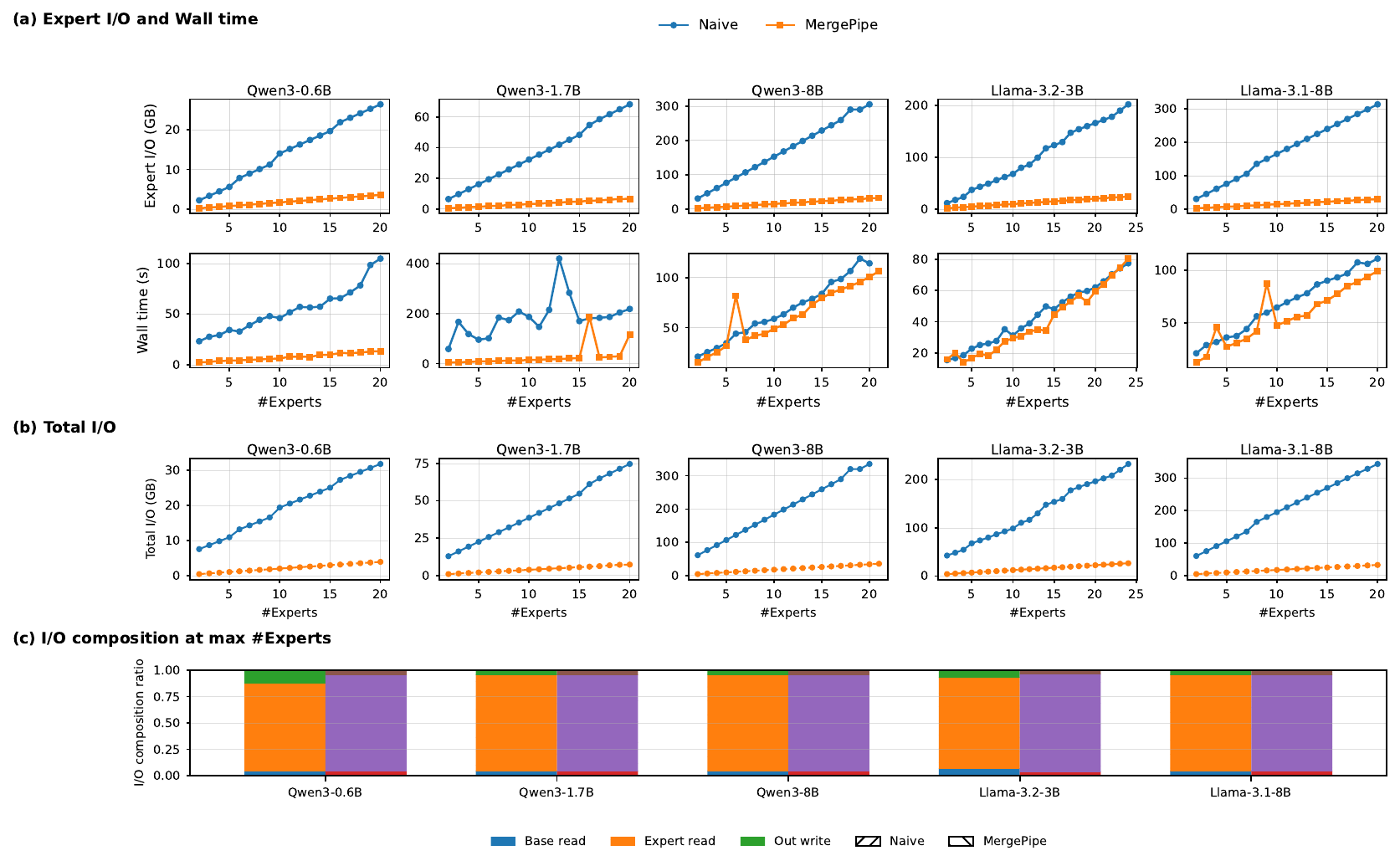}
  \caption{
    \textbf{Scaling with the number of experts.}
    \textbf{(a)} Expert read I/O (top) and end-to-end wall time (bottom).
    Naive merging repeatedly scans expert checkpoints, leading to near-linear growth in expert I/O and runtime as the number of experts increases.
    MergePipe enforces an explicit expert I/O budget at execution time, keeping expert reads bounded and significantly reducing wall time.
    \textbf{(b)} Total I/O shows that naive pipelines become increasingly expert-read dominated, while MergePipe reduces total I/O by limiting expert access.
    \textbf{(c)} I/O composition at the maximum number of experts illustrates that MergePipe shifts the dominant cost from expert reads to unavoidable base reads and output writes.
    All experiments are conducted on CPU using the same merge operator (TIES).
    }
  \label{fig:fig7_vldb_all}
\end{figure*}

\subsection{Experimental Setup}
\label{sec:exp-setup}

\textbf{Models and experts.}
We evaluate MergePipe on five representative open-source LLMs from the
Llama and Qwen families:
\{\textit{Llama-3.1-8B}, \textit{Llama-3.2-3B}, \textit{Qwen3-0.6B},
\textit{Qwen3-1.7B}, and \textit{Qwen3-8B}\}.
These models span parameter scales from 0.6B to 8B.
All experiments use the official HuggingFace checkpoints.
Table~\ref{tab:model-specs} summarizes the key model characteristics, including parameter count and approximate BF16 checkpoint size (estimated assuming 2 bytes per parameter).

For each base model, we construct merging workloads by progressively increasing the number of expert checkpoints.
Unless otherwise stated, experts are derived from the same base model and stored as full checkpoints or parameter deltas.

\noindent
\textbf{Merging operators.}
We evaluate MergePipe using two widely adopted sparse merging algorithms,
\textbf{TIES}~\citep{yadav2023ties} and \textbf{DARE}~\citep{yu2024language}.
Both methods selectively incorporate expert parameters and therefore
naturally benefit from budget-aware expert access.
Although MergePipe also supports average merging, our evaluation focuses
on TIES and DARE as they better reflect realistic large-scale merging
workflows.

\noindent
\textbf{Baselines.}
We compare MergePipe against naive merging pipelines that implement the
same merging operators but treat model parameters as unstructured files.
Each merge invocation independently scans all required expert parameters,
with no reuse, planning, or budget enforcement, reflecting the execution
model of existing open-source merging scripts.

\noindent
\textbf{Metrics.}
We report (i) \textbf{expert read I/O} (MB),
(ii) \textbf{end-to-end wall-clock time}, and
(iii) \textbf{relative improvement} over naive execution.
These metrics directly correspond to the cost components defined in
Section~\ref{sec:cost-model}, with a focus on the dominant expert read cost
$C_{\text{expert}}$.

\noindent
\textbf{System configuration.}
All experiments are conducted on a CPU-only server with two Intel Xeon
Platinum 8358 processors (64 physical cores, 128 hardware threads total)
and SSD-backed storage.
OS-level file caching is disabled unless otherwise stated.
No GPU acceleration is used, ensuring that merge performance is
I/O-bound rather than compute-bound.

\noindent
\textbf{Reproducibility.}
MergePipe persists merge plans, execution manifests, and lineage metadata,
enabling deterministic re-execution and post-hoc inspection.

\subsection{I1. Scaling with Number of Experts}
\label{sec:exp-scaling}

We study how MergePipe scales as the expert set grows, using TIES as the merge operator and varying the number of experts from 2 up to 20 (25 for Llama-3.2-3B). Figure~\ref{fig:fig7_vldb_all} summarizes expert I/O, end-to-end wall time, and total I/O across five model families.

\noindent
\textbf{Expert I/O scales from $O(K)$ to budgeted.}
Figure~\ref{fig:fig7_vldb_all}(a, top) shows that the naive pipeline exhibits near-linear growth in \emph{expert read} I/O as $K$ increases, matching the $O(K)$ term $C_{\text{expert}}$ in our cost model (Section~\ref{sec:cost-model}). In contrast, MergePipe keeps expert I/O \emph{nearly flat} across $K$ by enforcing the expert-read budget $C_{\text{expert}}\le B$, yielding substantial reductions at the maximum $K$: e.g., for Qwen3-8B and Llama-3.1-8B at $K{=}20$, naive expert I/O reaches hundreds of GB, while MergePipe remains in the tens of GB range (roughly an order-of-magnitude reduction).

\noindent
\textbf{Wall time follows expert I/O, with a few localized outliers.}
Figure~\ref{fig:fig7_vldb_all}(a, bottom) reports wall-clock time. Overall, wall time tracks expert I/O: naive merging slows down steadily as $K$ grows, while MergePipe remains significantly faster for all models and large $K$ (e.g., the largest gains appear for Qwen3-1.7B and Qwen3-8B where naive repeatedly scans large expert checkpoints).
We observe several \emph{localized} spikes that are not explained by the cost model (since they do not coincide with increased expert I/O):
(i) \textit{Qwen3-1.7B (naive)} shows a sharp wall-time spike around $K{\approx}13$--$14$;
(ii) \textit{Qwen3-8B (MergePipe)} shows an isolated spike around $K{\approx}6$;
(iii) \textit{Llama-3.1-8B (MergePipe)} shows a spike around $K{\approx}9$--$10$.
These outliers are consistent with transient \emph{system-level} disturbances (CPU scheduling and frequency scaling, background disk contention, and NUMA effects) rather than merge semantics: importantly, they occur without corresponding jumps in expert I/O and the curves immediately return to their prior trend. Even at these points, MergePipe remains well below the naive runtime at comparable $K$ for large models.

\noindent
\textbf{Total I/O becomes dominated by unavoidable costs.}
Figure~\ref{fig:fig7_vldb_all}(b) shows total I/O. Naive total I/O grows almost linearly with $K$ because expert reads dominate $C_{\text{merge}}$. With MergePipe, total I/O grows slowly and is increasingly dominated by the fixed costs $C_{\text{base}}$ and $C_{\text{out}}$.
This shift is made explicit in Figure~\ref{fig:fig7_vldb_all}(c): at maximum $K$, naive runs are largely \emph{expert-read dominated}, whereas MergePipe substantially reduces the expert-read fraction, making base reads and output materialization the primary contributors.

\begin{figure}[t]
  \centering
  \begin{minipage}[t]{0.48\linewidth}
    \centering
    \includegraphics[width=\linewidth]{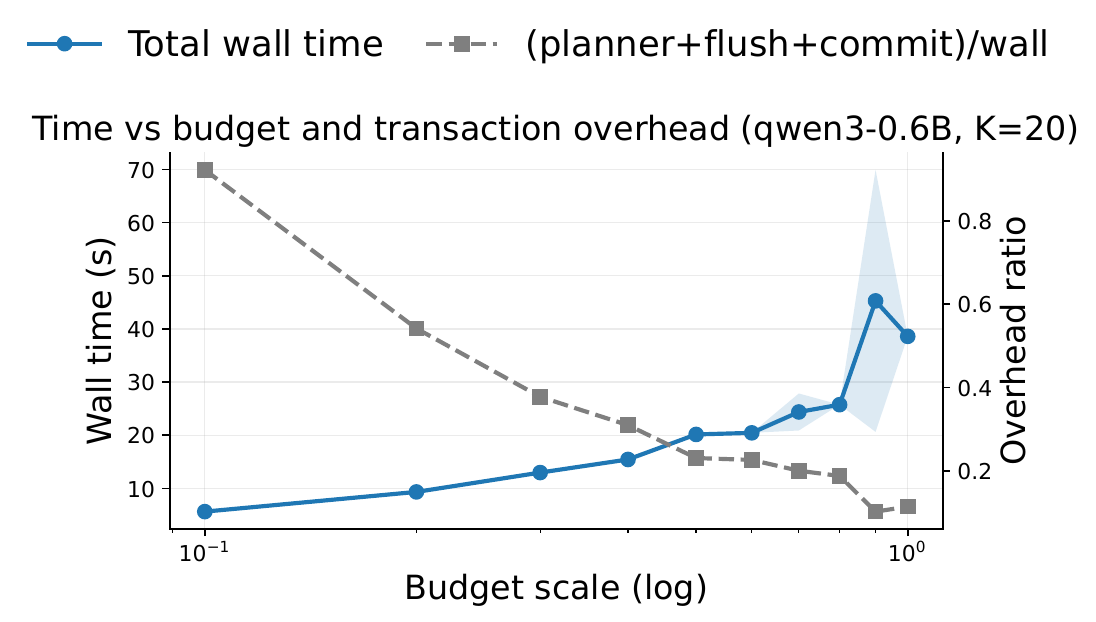}
  \end{minipage}
  \begin{minipage}[t]{0.48\linewidth}
    \centering
    \includegraphics[width=\linewidth]{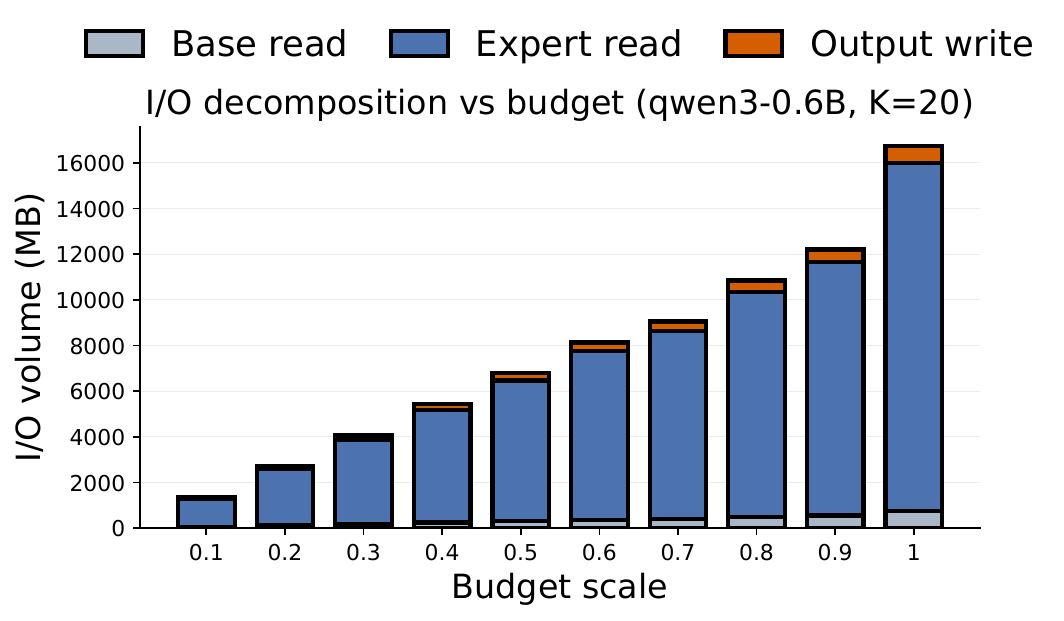}
  \end{minipage}
  \\
  \begin{minipage}[t]{0.95\linewidth}
    \centering
    \includegraphics[width=\linewidth]{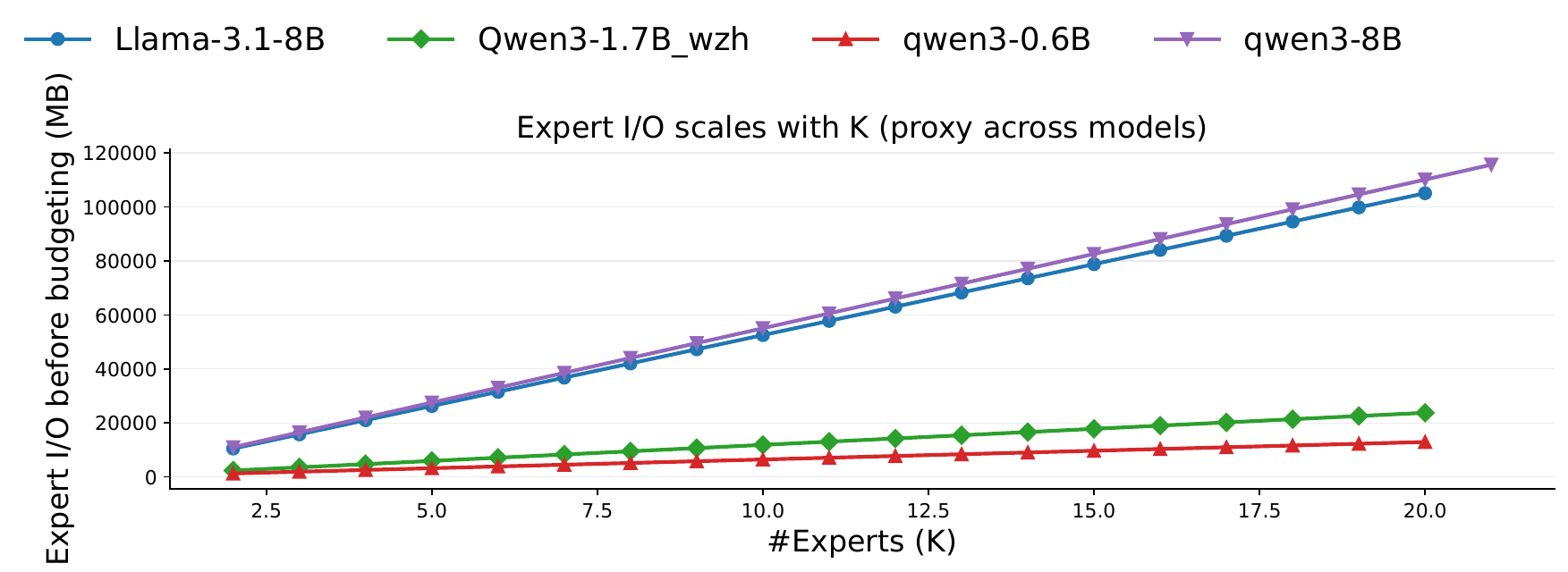}
  \end{minipage}
  
  \caption{
  {(a)} Planning+flush+commit is a small fraction of wall time. 
  {(b)} Budgeting mainly reduces expert reads.
  {(c)} Pre-budget expert read I/O scales with $K$ across models.
  }
\label{fig:cost-overhead}
\end{figure}

\noindent
\textbf{Where do the gains come from?}
To address concerns that improvements may stem from implementation details rather
than our execution model, we further decompose I/O and overheads.
Figure~\ref{fig:cost-overhead} shows that (i) pre-budget expert reads scale
approximately linearly with $K$ across models, (ii) varying the budget mainly
changes expert reads while base reads and output writes remain nearly constant,
and (iii) planning+flush+commit accounts for a small fraction of wall time over
practical budgets.
Together, these results indicate that MergePipe’s gains primarily arise from
reducing the dominant expert-read term rather than metadata or transactional
overheads.

\noindent
\textbf{Baseline strengthening.}
To reduce the risk that our gains come from an under-engineered baseline, we additionally report a disable-budget ablation under the same metric interface.
Table~\ref{tab:baseline-ablation} shows that MergePipe remains substantially better than naive full-read, and disabling planner-side budget scaling degrades runtime under the same I/O cap.

\begin{table}[t]
\centering
\small
\setlength{\tabcolsep}{5pt}
\caption{\textbf{Baseline \& disable-budget ablation} (Qwen3-0.6B, $K{=}20$, budget=8.57GB).
MergePipe’s gains persist over naive full-read; disabling planner-side budget scaling increases wall time under the same I/O cap.}
\label{tab:baseline-ablation}
\resizebox{\linewidth}{!}{
\begin{tabular}{lccc}
\toprule
\textbf{Method} & \textbf{Expert read (GB)} & \textbf{Total I/O (GB)} & \textbf{Wall (s)}\\
\midrule
Naive (full-read) & 26.4 & 31.8 & 104.7 \\
MergePipe (budgeted) & 8.0 & 8.8 & 22.8 \\
MergePipe (disable-budget) & 8.0 & 8.8 & 34.5 \\
\bottomrule
\end{tabular}}
\end{table}

As the number of experts increases, naive merging quickly becomes I/O-bound due to $O(K)$ expert scans. MergePipe enforces the budgeted expert-read term in Section~\ref{sec:cost-model}, achieving controlled scaling with substantially lower expert I/O and end-to-end time, while the remaining cost is increasingly limited by unavoidable base read and output write overheads.
Across all models and expert scales, the relative trends are consistent, indicating that the observed benefits stem from the execution model rather than model-specific effects.

\subsection{I2. Budget-Aware Planning Behavior}
\label{sec:exp-budget}

\begin{figure}[t]
  \centering
  \includegraphics[width=0.99\linewidth]{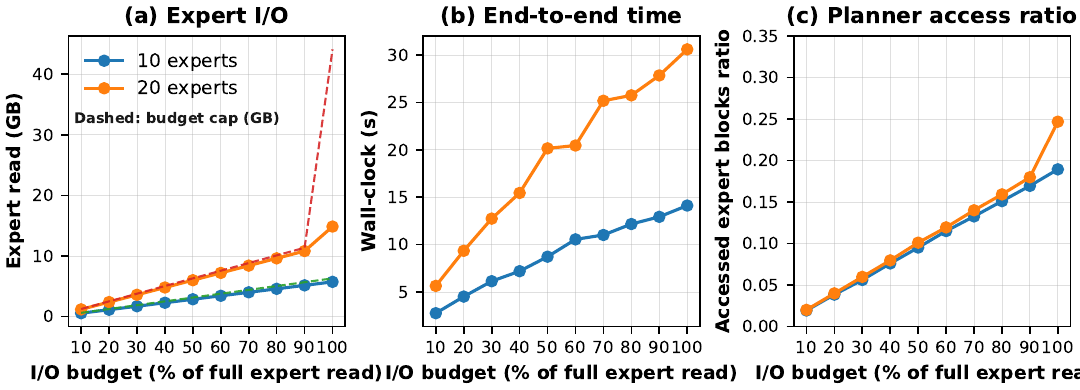}
  \caption{
\textbf{Budget-aware planning.}
\textbf{(a)} Expert read I/O grows monotonically with the I/O budget and is capped by the theoretical budget limit.
\textbf{(b)} End-to-end merge time closely follows expert I/O.
\textbf{(c)} Fraction of accessed expert blocks selected by the planner increases smoothly with the budget.
}
  \label{fig:fig7_4_vldb}
\end{figure}

We next study how MergePipe’s planner behaves under explicit I/O budget
constraints.
Fixing the base model, merge operator (TIES), and the number of experts,
we vary the expert-read budget from 10\% to 100\% of the full expert
parameter size.
Figure~\ref{fig:fig7_4_vldb} summarizes expert I/O, end-to-end latency,
and planner access behavior.

\noindent
\textbf{Expert I/O follows the imposed budget.}
Figure~\ref{fig:fig7_4_vldb}(a) shows the total expert read volume under
different budgets for workloads with 10 and 20 experts.
Actual expert I/O increases monotonically with the budget and always
remains below the theoretical budget cap (dashed lines), confirming that
MergePipe strictly enforces the user-specified constraint.
At the same budget level, workloads with more experts incur higher expert
I/O, indicating that planning decisions still reflect the underlying
problem scale rather than collapsing to a fixed heuristic.

\noindent
\textbf{End-to-end time tracks budget-controlled I/O.}
As shown in Figure~\ref{fig:fig7_4_vldb}(b), wall-clock time increases
monotonically with the I/O budget and closely follows the trend of expert
I/O.
This confirms that merging performance in this regime is I/O-bound, and
that adjusting the budget provides a direct and predictable handle on
end-to-end execution time.
We observe mild non-linearities around mid-range budgets (e.g., $\sim$50\%),
which stem from the discrete inclusion of additional expert blocks rather
than unstable system behavior; beyond these points, execution time
continues to grow smoothly.

\noindent
\textbf{Planner access expands in a controlled and interpretable manner.}
Figure~\ref{fig:fig7_4_vldb}(c) reports the fraction of expert parameter
blocks accessed by the planner.
The accessed-block ratio increases approximately linearly with the budget
and exhibits nearly identical trends for 10 and 20 experts.
This indicates that planner decisions are primarily budget-driven and
scale-invariant with respect to the total number of experts.
Together, these results demonstrate that MergePipe’s planner expands the
accessed expert set gradually and predictably as the budget increases,
providing fine-grained and user-controllable trade-offs between I/O cost
and execution time.

\subsection{I3. Generality Across Merge Operators}
\label{sec:exp-generality}

\begin{table}[t]
\centering
\Huge
\setlength{\tabcolsep}{2.5pt}
\renewcommand{\arraystretch}{1.4}
\caption{
\textbf{Generality across merge operators (Llama-3.1-8B).}
Comparison of naive execution and MergePipe under three merging operators.
\textbf{Top:} total parameter I/O (MB) and the fraction of naive I/O consumed by MergePipe (\texttt{Ratio}).
\textbf{Bottom:} end-to-end wall-clock time (seconds) and relative improvement
(\texttt{Improv.} = $(T_{\text{naive}}-T_{\text{MergePipe}})/T_{\text{naive}}$).
}
\label{tab:merge-generality}
\resizebox{\linewidth}{!}{
\begin{tabular}{@{} c| *{3}{c}| *{3}{c}| *{3}{c}| @{}}
\toprule
\multicolumn{10}{c}{\textbf{Total Expert I/O (MB)}} \\
\cmidrule(lr){1-10}
& \multicolumn{3}{c}{\textbf{Average}} &
  \multicolumn{3}{c}{\textbf{TIES}} &
  \multicolumn{3}{c}{\textbf{DARE}} \\
\cmidrule(lr){2-4} \cmidrule(lr){5-7} \cmidrule(lr){8-10}
$K$ & Naive & MergePipe & \textbf{Ratio} &
Naive & MergePipe & \textbf{Ratio} &
Naive & MergePipe & \textbf{Ratio} \\
\midrule
2 & 32127.62 & 3461.73 & \textbf{10.77\%} & 32127.62 & 3461.73 & \textbf{10.77\%} & 32127.62 & 3461.73 & \textbf{10.77\%} \\
4 & 44978.67 & 3461.73 & \textbf{7.70\%}  & 45766.68 & 3461.73 & \textbf{7.56\%}  & 44978.67 & 3461.73 & \textbf{7.70\%}  \\
6 & 57829.73 & 3461.73 & \textbf{5.99\%}  & 65831.24 & 3461.73 & \textbf{5.26\%}  & 65831.24 & 3461.73 & \textbf{5.26\%}  \\
8 & 70680.79 & 3461.73 & \textbf{4.90\%}  & 79470.30 & 3461.73 & \textbf{4.36\%}  & 79470.30 & 3461.73 & \textbf{4.36\%}  \\
10 & 83531.85 & 3461.73 & \textbf{4.14\%}  & 92321.36 & 3461.73 & \textbf{3.75\%}  & 92321.36 & 3461.73 & \textbf{3.75\%}  \\
12 & 102808.41& 3461.73 & \textbf{3.37\%}  & 105172.42& 3461.73 & \textbf{3.29\%}  & 105172.42& 3461.73 & \textbf{3.29\%}  \\
14 & 123660.97& 3461.73 & \textbf{2.80\%}  & 126024.99& 3461.73 & \textbf{2.75\%}  & 126024.99& 3461.73 & \textbf{2.75\%}  \\
16 & 137300.04& 3461.73 & \textbf{2.52\%}  & 140046.71& 3461.73 & \textbf{2.47\%}  & 139200.14& 3461.73 & \textbf{2.49\%}  \\
18 & 158152.60& 3461.73 & \textbf{2.19\%}  & 161316.03& 3461.73 & \textbf{2.15\%}  & 160200.60& 3461.73 & \textbf{2.16\%}  \\
20 & 171003.66& 3461.73 & \textbf{2.02\%}  & 174424.50& 3461.73 & \textbf{1.98\%}  & 173200.17& 3461.73 & \textbf{2.00\%}  \\
\midrule
\multicolumn{10}{c}{\textbf{End-to-End Wall Time (seconds)}} \\
\cmidrule(lr){1-10}
& \multicolumn{3}{c}{\textbf{Average}} &
  \multicolumn{3}{c}{\textbf{TIES}} &
  \multicolumn{3}{c}{\textbf{DARE}} \\
\cmidrule(lr){2-4} \cmidrule(lr){5-7} \cmidrule(lr){8-10}
$K$ & Naive & MergePipe & \textbf{Improv.} &
Naive & MergePipe & \textbf{Improv.} &
Naive & MergePipe & \textbf{Improv.} \\
\midrule
2 & 31.11 & 41.17 & \textbf{-32.34\%} & 404.27 & 37.17 & \textbf{90.81\%} & 79.27 & 37.78 & \textbf{52.34\%} \\
4 & 37.31 & 50.98 & \textbf{-36.64\%} & 491.77 & 42.31 & \textbf{91.40\%} & 78.14 & 45.40 & \textbf{41.90\%} \\
6 & 81.93 & 56.69 & \textbf{30.81\%} & 558.78 & 46.32 & \textbf{91.71\%} & 92.07 & 43.59 & \textbf{52.66\%} \\
8 & 136.17 & 64.07 & \textbf{52.95\%} & 614.27 & 51.06 & \textbf{91.69\%} & 99.10 & 44.72 & \textbf{54.87\%} \\
10 & 159.95 & 77.36 & \textbf{51.63\%} & 578.84 & 62.89 & \textbf{89.14\%} & 101.03 & 67.22 & \textbf{33.47\%} \\
12 & 191.94 & 84.22 & \textbf{56.12\%} & 633.06 & 80.28 & \textbf{87.32\%} & 127.99 & 84.73 & \textbf{33.80\%} \\
14 & 172.67 & 127.30 & \textbf{26.28\%} & 790.65 & 90.97 & \textbf{88.49\%} & 206.89 & 153.13 & \textbf{25.98\%} \\
16 & 195.17 & 173.52 & \textbf{11.09\%} & 941.36 & 198.65 & \textbf{78.90\%} & 248.91 & 210.03 & \textbf{15.62\%} \\
18 & 361.59 & 334.47 & \textbf{7.50\%} & 1051.48& 314.52 & \textbf{70.09\%} & 395.30 & 358.85 & \textbf{9.22\%} \\
20 & 363.02 & 338.36 & \textbf{6.79\%} & 1203.65& 356.32 & \textbf{70.40\%} & 464.16 & 414.44 & \textbf{10.71\%} \\
\bottomrule
\end{tabular}
}
\end{table}

We evaluate the generality of MergePipe across different merging algorithms.
Table~\ref{tab:merge-generality} compares naive execution and MergePipe under
three representative operators: Average, TIES, and DARE, while varying the
number of experts $K$ from 2 to 20.

\noindent
\textbf{Consistent expert I/O control across operators.}
Across all three operators and all expert set sizes, MergePipe reduces total
parameter I/O by nearly an order of magnitude.
While naive execution exhibits near-linear growth in total I/O as $K$
increases (e.g., exceeding 170\,GB at $K{=}20$), MergePipe keeps total I/O
almost constant at approximately 3.5\,GB.
As a result, the I/O ratio (\texttt{Ratio}) steadily decreases with $K$,
dropping to about 2\% at $K{=}20$ for all operators.
This confirms that MergePipe’s I/O savings are independent of merge semantics
and arise from budget-aware parameter access rather than operator-specific
optimizations.

\noindent
\textbf{Runtime benefits depend on merge sparsity.}
End-to-end runtime improvements vary across operators.
For sparse merging algorithms such as TIES and DARE, MergePipe achieves
substantial speedups across all $K$.
For example, under TIES, MergePipe reduces wall time by over $90\%$ at
$K{=}4$--$8$ and still delivers about $70\%$ improvement at $K{=}20$.
These gains occur because naive sparse merging repeatedly scans large expert
checkpoints, whereas MergePipe selectively accesses a budgeted subset of
parameters.

In contrast, Average merging shows smaller and occasionally negative
improvements at very small $K$ (e.g., $K{=}2$ and $4$), where naive execution
is already fast and planning overhead is not yet amortized.
However, as $K$ increases, MergePipe consistently outperforms naive execution,
achieving over $50\%$ runtime reduction at moderate scales (e.g., $K{=}8$--$12$).

\noindent
\textbf{Case analysis.}
At $K{=}8$, MergePipe reduces total I/O from 70--80\,GB to 3.5\,GB across all
operators, while reducing wall time from 614\,s to 51\,s under TIES and from
99\,s to 45\,s under DARE.
At $K{=}20$, despite different merge semantics, all operators exhibit similar
I/O ratios (around 4\%), indicating that MergePipe’s planner enforces the same
budgeted access pattern regardless of how expert parameters are combined.
Overall, these results demonstrate that MergePipe generalizes across merge operators, consistently controls the dominant expert I/O cost, and delivers the largest performance benefits for realistic sparse merging workloads.

\subsection{I4. Execution Overheads and System Costs}
\label{sec:exp-overheads}

In this subsection, we want to answer the question "Does MergePipe introduce non-negligible system overheads from planning, metadata management, and transactional execution compared with end-to-end merging?"
We evaluate whether these components incur meaningful overheads in practice.

\begin{table}[t]
\centering
\caption{
Execution overheads and system costs of MergePipe under a representative configuration (Qwen3-0.6B, 16 experts).
Planning and metadata overheads are small relative to execution time and total I/O.
}
\label{tab:execution_overheads}
\resizebox{\linewidth}{!}{
\begin{tabular}{lcc}
\toprule
\textbf{Metric} & \textbf{Absolute Value} & \textbf{Relative Cost} \\
\midrule
Planning time ($T_{\text{plan}}$) & 1.21 s & 1.04\% of execution \\
Execution time ($T_{\text{exec}}$) & 116.62 s & 100\% \\
\midrule
Estimated expert I/O (pre-plan) & 19\,842 MB & -- \\
Executed expert I/O & 23\,349 MB & 1.18$\times$ \\
Total I/O volume & 91\,447 MB & 100\% \\
\midrule
Metadata catalog size (DB) & 3\,468 MB & 3.79\% of total I/O \\
Execution manifest size & 812 KB & $<$0.001\% \\
\bottomrule
\end{tabular}}
\end{table}

\noindent
\textbf{Planning overhead is negligible.}
As shown in Table~\ref{tab:execution_overheads}, planning incurs only 1.21\,s, accounting for 1.04\% of the end-to-end execution time.
This confirms that MergePipe’s cost modeling, conflict resolution, and plan generation introduce minimal overhead compared with merge execution.
Across different expert counts, we observe similarly small planning overheads, indicating that planner cost does not scale with the number of experts.

\noindent
\textbf{Planner-induced I/O is amortized by execution savings.}
Conflict aware planning may introduce additional expert reads to validate candidate accesses.
In this configuration, executed expert I/O increases by 1.18$\times$ relative to the pre-planning estimate.
However, this modest overhead is amortized by eliminating redundant expert scans during execution, resulting in substantially lower total I/O compared with naive merging.

\noindent
\textbf{Metadata and transactional costs are modest.}
Persistent metadata, including the SQLite catalog and lineage information, accounts for only 3.79\% of total I/O volume, while execution manifests are negligible in size.
These costs remain small compared with model parameter I/O and do not affect end-to-end merge throughput.
These results show that MergePipe introduces only minor system overheads, while effectively reducing dominant expert I/O costs.
Planning, metadata management, and transactional execution scale favorably and do not offset the performance benefits of budget-aware merging.

\subsection{I5. Ablation and Sensitivity Analysis}
\label{sec:exp-ablation}

\begin{table}[t]
  \centering
  \caption{
Sensitivity to block size in MergePipe under fixed expert set and I/O budget.
Very small blocks increase overhead, while overly large blocks reduce budget control granularity.
}

  \label{tab:block-size-sensitivity}
  \vspace{2pt}
  \setlength{\tabcolsep}{5.2pt}
  \renewcommand{\arraystretch}{1.15}
  \resizebox{\linewidth}{!}{
  \begin{tabular}{l l r r r r r r}
    \toprule
    \multicolumn{2}{c}{\textbf{Block size}} &
    \textbf{16\,k} & \textbf{32\,k} & \textbf{64\,k} &
    \textbf{128\,k} & \textbf{256\,k} & \textbf{512\,k} \\
    \midrule

    \multirow{2}{*}{\textbf{TIES}} &
    Expert I/O (MB)  & 4039.55 & 3520.03 & 3226.09 & 3198.74 & 3390.15 & 3772.96 \\
    & Wall time (s)  & 77.33   & 61.50   & 52.00   & 45.32   & 46.33   & 45.04 \\
    \midrule

    \multirow{2}{*}{\textbf{DARE}} &
    Expert I/O (MB)  & 4039.55 & 3520.03 & 3226.09 & 3198.74 & 3390.15 & 3772.96 \\
    & Wall time (s)  & 83.89   & 62.47   & 50.38   & 46.31   & 46.33   & 46.82 \\
    \bottomrule
  \end{tabular}}
\end{table}

In this subsection, we want to answer "Which design choices are critical for MergePipe’s I/O reduction, and how sensitive is performance to key system parameters?" We study the sensitivity of MergePipe to the block granularity used during planning and execution.
Block size determines the atomic unit of parameter access and directly affects both budget controllability and execution overhead.

\noindent
\textbf{Sensitivity to block size.}
Table~\ref{tab:block-size-sensitivity} reports total expert I/O volume and end-to-end wall-clock time under varying block sizes for both TIES and DARE operators.
Across both operators, performance exhibits a clear and consistent trend.
When blocks are too small, like 16\,k or 32\,k), total I/O and execution time increase noticeably.
This is due to over-fragmentation: the planner must manage and access a large number of fine-grained blocks, increasing metadata lookups, index traversal, and I/O seek overhead.
As block size increases, both expert I/O and wall-clock time decrease, reaching their minimum around 64\,k--128\,k.
In this regime, blocks are sufficiently fine-grained to allow effective budget-aware selection, while avoiding excessive fragmentation.
For overly large blocks (256\,k and above), total I/O begins to rise again.
Here, budget control becomes coarse: selecting a single large block may pull in unnecessary parameters, reducing the effectiveness of selective expert access.
Despite this, wall-clock time remains relatively stable, indicating that MergePipe degrades gracefully rather than exhibiting pathological behavior.

\noindent
\textbf{Robustness across merge operators.}
Notably, TIES and DARE exhibit nearly identical sensitivity curves across all block sizes.
This suggests that block-size sensitivity is primarily a system-level phenomenon, independent of merge semantics. MergePipe therefore provides a robust operating regime with respect to block granularity, without requiring operator-specific tuning.
These results show that while block size influences the efficiency of MergePipe, the system is not overly sensitive to this parameter.
A broad range of block sizes yields near-optimal performance, validating the robustness of block-level planning in practical deployments.

\begin{figure}[t]
  \centering
  \includegraphics[width=\linewidth]{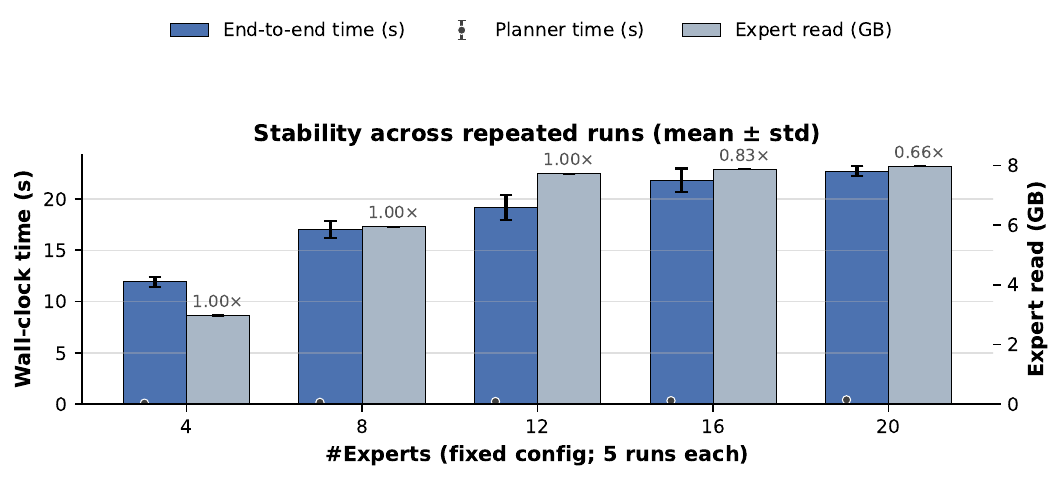}
  \caption{
Stability across repeated runs.
Mean and standard deviation over five identical executions under a fixed configuration (Qwen3-0.6B, TIES).
End-to-end time and expert I/O exhibit low variance, indicating stable execution behavior.
}

  \label{fig:stability}
\end{figure}

\subsection{Stability Across Repeated Runs}
\label{sec:exp-stability}

To assess the stability and repeatability of MergePipe under identical configurations, we conduct a repeated-run experiment by fixing the model, merge operator, I/O budget, and planner configuration, and varying only the number of experts.
For each expert count $K \in \{4, 8, 12, 16, 20\}$, we execute the same merge workload \textbf{five times} and report the mean and standard deviation of key metrics.

The overall results indicate that \textbf{MergePipe’s execution behavior is deterministic} given a fixed plan, and that observed variance is dominated by system-level noise rather than planning instability.
Figure~\ref{fig:stability} summarizes the results. End-to-end wall-clock time exhibits a smooth and monotonic increase as the number of experts grows, from approximately $12$\,s at $K{=}4$ to $22$--$23$\,s at $K{=}20$.
Across all settings, the observed standard deviation is small (typically below $5\%$ of the mean), indicating that end-to-end performance is highly stable under repeated execution. Even at larger expert counts, where absolute runtime is higher, variance remains limited like $\pm 1$\,s around a $21$--$22$\,s mean at $K{=}16$.

Expert read volume is even more stable.
As shown by the right axis in Figure~\ref{fig:stability}, the amount of expert parameters accessed under a fixed I/O budget is nearly deterministic, with negligible variance across runs.
For example, at $K{=}12$, $16$, and $20$, the mean expert read remains close to $1.00\times$, $0.83\times$, and $0.66\times$ of the full expert read, respectively, with no visible fluctuation across repetitions. This confirms that MergePipe’s planner makes consistent block-selection decisions under identical inputs, and that expert I/O variability does not contribute to runtime variance.

Planner overhead is both small and stable. It is noticeable that Planner execution time remains below $0.2$\,s across all configurations, with very small standard deviation. Consequently, planner variability is negligible compared to the overall execution time and does not affect end-to-end stability.
It demonstrates that MergePipe delivers \emph{stable and repeatable} performance under repeated runs. The small runtime variance observed in practice is attributable to normal system-level effects such as CPU scheduling and background I/O rather than to unstable planning behavior or inconsistent expert access. All experimental results in this section are reported as averages over five runs.

\begin{table}[t]
\centering
\large
\caption{Parameter deviation and downstream performance under different budgets
(Qwen3-0.6B, TIES, $K{=}20$).}
\label{tab:correctness_quality}
\setlength{\tabcolsep}{4pt}
\resizebox{\linewidth}{!}{
\begin{tabular}{c|ccc|ccc}
\toprule
 & \multicolumn{3}{c|}{\textbf{Deviation}} 
 & \multicolumn{3}{c}{\textbf{Benchmarks}} \\
\cmidrule(lr){2-4} \cmidrule(lr){5-7}
\textbf{Budget} 
& Touched Ratio 
& Rel. $\ell_2$ Err. 
& P95 Block Err. 
& HumanEval 
& IFEval 
& DROP \\
\midrule
1.0 (Full) & 0.247 & 0 & 0 & 39.71 & 68.56 & 66.81 \\
0.9 & 0.180 & 7.23e-4 & 3.66e-3 & 42.68 & 68.82 & 67.23 \\
0.8 & 0.159 & 7.77e-4 & 3.66e-3 & 36.59 & 68.94 & 67.08 \\
0.7 & 0.140 & 8.30e-4 & 3.98e-3 & 37.80 & 68.47 & 66.26 \\
0.6 & 0.119 & 8.30e-4 & 3.98e-3 & 37.37 & 67.87 & 66.10 \\
0.5 & 0.101 & 8.84e-4 & 3.98e-3 & 40.24 & 68.11 & 66.00 \\
\bottomrule
\end{tabular}}
\end{table}

\subsection{Correctness and Quality Preservation}
\label{sec:exp-correctness}

We assess whether budget-aware execution preserves the correctness and quality of the merged models.
MergePipe does not change the merge performance; it only constrains expert parameter access under a budget.

\noindent
\textbf{(A) Parameter-level deviation.}
Table~\ref{tab:correctness_quality} compares the budgeted output $\theta_B$ against the full-read output $\theta_{\text{full}}$ for the same setting (Qwen3-0.6B, TIES, $K{=}20$), reporting the relative $\ell_2$ error $\|\theta_B-\theta_{\text{full}}\|_2 / \|\theta_{\text{full}}\|_2$ and a high-percentile block error.
Across budgets down to $B{=}0.5$, deviations remain small (on the order of $10^{-3}$), and increase smoothly as fewer expert blocks are accessed (Touched Ratio), indicating controlled approximation.

\noindent
\textbf{(B) Downstream quality.}
Table~\ref{tab:correctness_quality} reports task performance on three benchmarks covering code generation (HumanEval \citep{chen2021evaluating}), instruction following (IFEval \citep{zhou2023instruction}), and text reasoning (DROP \citep{dua2019drop}).
Scores remain close to the full-read baseline across all budgets, with no consistent degradation as the budget decreases.
Together, the parameter-level and task-level results suggest that MergePipe achieves substantial expert I/O reduction while preserving merged model quality in this representative configuration.

\section{Related Work}
\label{sec:related_work}

\textbf{LLM merging} aims to consolidate multiple trained models into a single model without retraining from scratch and has become a common component in modern LLM development pipelines.
Existing approaches can be broadly categorized into \emph{dense} and \emph{sparse} merging operators.
Dense methods, such as weight averaging, model soups~\citep{wortsman2022model}, and Task Arithmetic (TA)~\citep{ilharcoediting}, combine parameters globally and implicitly assume that conflicts between task updates are resolved through averaging.
Sparse merging methods instead selectively incorporate task-specific parameters to reduce interference, including TIES~\citep{yadav2023ties}, DARE~\citep{yu2024language}, and SVD-based operators that exploit low-rank structures in task vectors~\citep{lu2024twin,liu2025lore}.
More recent activation-aware methods dynamically weight models based on input activations~\citep{nobari2025activation,liu2025sens}, while MoE-style approaches introduce routing mechanisms to combine experts at inference time~\citep{sukhbaatar2024branch,liu20251bit}.
Several studies further investigate empirical scaling laws and theoretical properties of model merging~\citep{wang2025model}, and comprehensive surveys summarize this growing literature~\citep{zhou2025democratizing}.

Despite their algorithmic diversity, existing merging methods predominantly focus on \emph{how} parameters should be combined, while implicitly assuming that expert parameters can be accessed at negligible cost.
In practical LLM pipelines, however, repeatedly reading expert checkpoints incurs substantial parameter I/O overhead that scales linearly with the number of experts.
This cost is both dominant and controllable, yet is largely abstracted away in prior work.
MergePipe addresses this orthogonal bottleneck by providing a budget-aware execution engine that enforces bounded expert reads, enabling large-scale merging without altering merge semantics.

\noindent
\textbf{LLM fusion} represents an alternative paradigm that aligns model behaviors through distillation or hybrid pipelines rather than direct parameter manipulation.
Prior work studies alignment \citep{wang2024interdependency} at the level of logits, representations, or architectural components~\citep{wang2025infigfusion}, primarily focusing on algorithmic design and quality transfer across models.
These approaches are complementary to MergePipe: while fusion methods aim to improve model performance, MergePipe focuses on making parameter-level merging \emph{executable, predictable, and cost-efficient} under explicit resource constraints, which is critical for iterative expert merge in large-scale LLM systems.

\noindent
\textbf{Machine Learning Management Systems.}
From a systems perspective, large-scale LLM development increasingly resembles a data-management workload over large immutable artifacts like checkpoints, deltas, and derived variants, where indexing, reuse, and provenance become first-class requirements.
Prior systems have studied model- and experiment-management abstractions like ModelDB for model lifecycle tracking and queryable metadata~\citep{hines2004modeldb, vartak2016modeldb}, as well as end-to-end ML platforms that record runs, parameters, and artifacts such as MLflow to improve reproducibility and operational visibility~\citep{zaharia2018accelerating}.
In parallel, data and version control tools for ML pipelines like DVC \citep{barreto2024dvc} and other works \citep{van2017versioning,vadde2024devops} emphasize dataset and artifact versioning, repeatable pipelines, and auditable diffs.
Workflow engines and pipeline systems \citep{eggers2024automating,bux2015saasfee} provide orchestration and caching, but typically do not model \emph{parameter-access cost} as an explicit optimization objective.
More broadly, database research on lineage and provenance has established principled foundations and practical systems for tracking derivations and enabling post-hoc inspection in data workflows~\citep{buneman2001and, green2007provenance}.

In contrast, MergePipe is complementary to prior provenance and pipeline systems, but focuses on a different system-level bottleneck.
Existing provenance systems~\citep{ruan2021lineagechain} typically treat model checkpoints as opaque artifacts, and therefore lack visibility into the cost structure of iterative LLM merging.
As a result, they cannot reason about or control the dominant I/O pattern in such workloads, where repeatedly scanning expert checkpoints induces an $\mathcal{O}(K)$ expert-read cost.
MergePipe instead treats model parameters as manageable data.
It (i) organizes checkpoints into tensors or blocks with persistent statistics, (ii) represents merge plans as reusable execution objects, and (iii) explicitly models expert parameter reads as a budgeted resource during planning and execution.
This budget-aware, block-granular execution model distinguishes MergePipe from prior pipeline and provenance tooling, which primarily improves traceability but does not provide bounded or predictable expert I/O behavior for large-scale merging.

\section{Conclusion}
\label{sec:conclusion}
We presented \textbf{MergePipe}, a budget-aware execution pipeline that reframes large-scale LLM merging as a data management and execution problem. By treating \emph{expert parameter reads} as the dominant and controllable cost, MergePipe uses catalog-driven planning and streaming execution to enforce explicit expert I/O budgets while remaining agnostic to merge operators. Across representative merging workloads, MergePipe cuts total I/O by up to an order of magnitude and achieves up to $11\times$ end-to-end speedups (up to 70\% wall-time reduction). 
MergePipe is most beneficial for offline merges with many experts where expert I/O dominates, while its gains naturally diminish for small expert sets or dense merges.



\section*{The Use of Large Language Models}
In this manuscript, we employed an LLM exclusively as a writing assistant, including correction of grammatical errors, rephrasing for better readability, and minor stylistic improvements. All substantive analysis, experiments, and conclusions are the original work of the authors. This usage complies with prevailing academic standards on the responsible application of generative AI tools.


\bibliographystyle{ACM-Reference-Format}
\bibliography{ref}

@article{naveed2025comprehensive,
  title={A comprehensive overview of large language models},
  author={Naveed, Humza and Khan, Asad Ullah and Qiu, Shi and Saqib, Muhammad and Anwar, Saeed and Usman, Muhammad and Akhtar, Naveed and Barnes, Nick and Mian, Ajmal},
  journal={ACM Transactions on Intelligent Systems and Technology},
  volume={16},
  number={5},
  pages={1--72},
  year={2025},
  publisher={ACM New York, NY}
}

@article{shao2024deepseekmath,
  title={Deepseekmath: Pushing the limits of mathematical reasoning in open language models},
  author={Shao, Zhihong and Wang, Peiyi and Zhu, Qihao and Xu, Runxin and Song, Junxiao and Bi, Xiao and Zhang, Haowei and Zhang, Mingchuan and Li, YK and Wu, Yang and others},
  journal={arXiv preprint arXiv:2402.03300},
  year={2024}
}

@article{yang2024qwen2,
  title={Qwen2. 5-math technical report: Toward mathematical expert model via self-improvement},
  author={Yang, An and Zhang, Beichen and Hui, Binyuan and Gao, Bofei and Yu, Bowen and Li, Chengpeng and Liu, Dayiheng and Tu, Jianhong and Zhou, Jingren and Lin, Junyang and others},
  journal={arXiv preprint arXiv:2409.12122},
  year={2024}
}

@article{hui2024qwen2,
  title={Qwen2. 5-coder technical report},
  author={Hui, Binyuan and Yang, Jian and Cui, Zeyu and Yang, Jiaxi and Liu, Dayiheng and Zhang, Lei and Liu, Tianyu and Zhang, Jiajun and Yu, Bowen and Lu, Keming and others},
  journal={arXiv preprint arXiv:2409.12186},
  year={2024}
}

@article{guo2024deepseek,
  title={DeepSeek-Coder: When the Large Language Model Meets Programming--The Rise of Code Intelligence},
  author={Guo, Daya and Zhu, Qihao and Yang, Dejian and Xie, Zhenda and Dong, Kai and Zhang, Wentao and Chen, Guanting and Bi, Xiao and Wu, Yu and Li, YK and others},
  journal={arXiv preprint arXiv:2401.14196},
  year={2024}
}

@inproceedings{yue2024lawllm,
  title={Lawllm: Intelligent legal system with legal reasoning and verifiable retrieval},
  author={Yue, Shengbin and Liu, Shujun and Zhou, Yuxuan and Shen, Chenchen and Wang, Siyuan and Xiao, Yao and Li, Bingxuan and Song, Yun and Shen, Xiaoyu and Chen, Wei and others},
  booktitle={International Conference on Database Systems for Advanced Applications},
  pages={304--321},
  year={2024},
  organization={Springer}
}

@article{cui2023chatlaw,
  title={Chatlaw: A multi-agent collaborative legal assistant with knowledge graph enhanced mixture-of-experts large language model},
  author={Cui, Jiaxi and Ning, Munan and Li, Zongjian and Chen, Bohua and Yan, Yang and Li, Hao and Ling, Bin and Tian, Yonghong and Yuan, Li},
  journal={arXiv preprint arXiv:2306.16092},
  year={2023}
}

@article{xu2025lingshu,
  title={Lingshu: A Generalist Foundation Model for Unified Multimodal Medical Understanding and Reasoning},
  author={Xu, Weiwen and Chan, Hou Pong and Li, Long and Aljunied, Mahani and Yuan, Ruifeng and Wang, Jianyu and Xiao, Chenghao and Chen, Guizhen and Liu, Chaoqun and Li, Zhaodonghui and others},
  journal={arXiv preprint arXiv:2506.07044},
  year={2025}
}

@article{yang2025qwen3,
  title={Qwen3 technical report},
  author={Yang, An and Li, Anfeng and Yang, Baosong and Zhang, Beichen and Hui, Binyuan and Zheng, Bo and Yu, Bowen and Gao, Chang and Huang, Chengen and Lv, Chenxu and others},
  journal={arXiv preprint arXiv:2505.09388},
  year={2025}
}

@article{agarwal2025gpt,
  title={gpt-oss-120b \& gpt-oss-20b model card},
  author={Agarwal, Sandhini and Ahmad, Lama and Ai, Jason and Altman, Sam and Applebaum, Andy and Arbus, Edwin and Arora, Rahul K and Bai, Yu and Baker, Bowen and Bao, Haiming and others},
  journal={arXiv preprint arXiv:2508.10925},
  year={2025}
}

@article{yang2024model,
  title={Model merging in llms, mllms, and beyond: Methods, theories, applications and opportunities},
  author={Yang, Enneng and Shen, Li and Guo, Guibing and Wang, Xingwei and Cao, Xiaochun and Zhang, Jie and Tao, Dacheng},
  journal={arXiv preprint arXiv:2408.07666},
  year={2024}
}

@article{wang2025model,
  title={Model Merging Scaling Laws in Large Language Models},
  author={Wang, Yuanyi and Gu, Yanggan and Zhang, Yiming and Zhou, Qi and Yan, Zhaoyi and Xie, Congkai and Wang, Xinyao and Yuan, Jianbo and Yang, Hongxia},
  journal={arXiv preprint arXiv:2509.24244},
  year={2025}
}

@inproceedings{sung2023empirical,
  title={An Empirical Study of Multimodal Model Merging},
  author={Sung, Yi-Lin and Li, Linjie and Lin, Kevin and Gan, Zhe and Bansal, Mohit and Wang, Lijuan},
  booktitle={Findings of the Association for Computational Linguistics: EMNLP 2023},
  pages={1563--1575},
  year={2023}
}

@article{lu2024twin,
  title={Twin-merging: Dynamic integration of modular expertise in model merging},
  author={Lu, Zhenyi and Fan, Chenghao and Wei, Wei and Qu, Xiaoye and Chen, Dangyang and Cheng, Yu},
  journal={Advances in Neural Information Processing Systems},
  volume={37},
  pages={78905--78935},
  year={2024}
}

@article{bidermanlora,
  title={LoRA Learns Less and Forgets Less},
  author={Biderman, Dan and Portes, Jacob and Ortiz, Jose Javier Gonzalez and Paul, Mansheej and Greengard, Philip and Jennings, Connor and King, Daniel and Havens, Sam and Chiley, Vitaliy and Frankle, Jonathan and others},
  journal={Transactions on Machine Learning Research}
}

@article{ganaie2022ensemble,
  title={Ensemble deep learning: A review},
  author={Hu, Minghui},
  journal={Engineering Applications of Artificial Intelligence},
  volume={115},
  pages={105151},
  year={2022},
  publisher={Elsevier}
}

@inproceedings{wortsman2022model,
  title={Model soups: averaging weights of multiple fine-tuned models improves accuracy without increasing inference time},
  author={Wortsman, Mitchell and Ilharco, Gabriel and Gadre, Samir Ya and Roelofs, Rebecca and Gontijo-Lopes, Raphael and Morcos, Ari S and Namkoong, Hongseok and Farhadi, Ali and Carmon, Yair and Kornblith, Simon and others},
  booktitle={International conference on machine learning},
  pages={23965--23998},
  year={2022},
  organization={PMLR}
}

@article{yadav2023ties,
  title={Ties-merging: Resolving interference when merging models},
  author={Yadav, Prateek and Tam, Derek and Choshen, Leshem and Raffel, Colin A and Bansal, Mohit},
  journal={Advances in Neural Information Processing Systems},
  volume={36},
  pages={7093--7115},
  year={2023}
}

@inproceedings{yu2024language,
  title={Language models are super mario: Absorbing abilities from homologous models as a free lunch},
  author={Yu, Le and Yu, Bowen and Yu, Haiyang and Huang, Fei and Li, Yongbin},
  booktitle={Forty-first International Conference on Machine Learning},
  year={2024}
}

@article{liu2024deepseek,
  title={Deepseek-v3 technical report},
  author={Liu, Aixin and Feng, Bei and Xue, Bing and Wang, Bingxuan and Wu, Bochao and Lu, Chengda and Zhao, Chenggang and Deng, Chengqi and Zhang, Chenyu and Ruan, Chong and others},
  journal={arXiv preprint arXiv:2412.19437},
  year={2024}
}

@article{lu2024merge,
  title={Merge, ensemble, and cooperate! a survey on collaborative strategies in the era of large language models},
  author={Lu, Jinliang and Pang, Ziliang and Xiao, Min and Zhu, Yaochen and Xia, Rui and Zhang, Jiajun},
  journal={arXiv preprint arXiv:2407.06089},
  year={2024}
}

@inproceedings{ilharcoediting,
  title={Editing models with task arithmetic},
  author={Ilharco, Gabriel and Ribeiro, Marco Tulio and Wortsman, Mitchell and Schmidt, Ludwig and Hajishirzi, Hannaneh and Farhadi, Ali},
  booktitle={The Eleventh International Conference on Learning Representations}
}

@inproceedings{wang2024interdependency,
  title={Interdependency matters: graph alignment for multivariate time series anomaly detection},
  author={Wang, Yuanyi and Sun, Haifeng and Wang, Chengsen and Zhu, Mengde and Wang, Jingyu and Tang, Wei and Qi, Qi and Zhuang, Zirui and Liao, Jianxin},
  booktitle={2024 IEEE International Conference on Data Mining (ICDM)},
  pages={869--874},
  year={2024},
  organization={IEEE}
}

@article{wang2025infigfusion,
  title={InfiGFusion: Graph-on-Logits Distillation via Efficient Gromov-Wasserstein for Model Fusion},
  author={Wang, Yuanyi and Yan, Zhaoyi and Zhang, Yiming and Zhou, Qi and Gu, Yanggan and Wu, Fei and Yang, Hongxia},
  journal={arXiv preprint arXiv:2505.13893},
  year={2025}
}

@article{zhou2025democratizing,
  title={Democratizing AI through model fusion: A comprehensive review and future directions},
  author={Zhou, Qi and Zhang, Yiming and Gu, Yanggan and Wang, Yuanyi and Sang, Zhijie and Yan, Zhaoyi and Li, Zhen and Zhang, Shengyu and Wu, Fei and Yang, Hongxia},
  journal={Nexus},
  year={2025},
  publisher={Elsevier}
}

@article{liu2025lore,
  title={LoRE-Merging: Exploring Low-Rank Estimation For Large Language Model Merging},
  author={Liu, Zehua and Wu, Han and Yao, Yuxuan and She, Ruifeng and Han, Xiongwei and Zhong, Tao and Yuan, Mingxuan},
  journal={arXiv preprint arXiv:2502.10749},
  year={2025}
}

@article{nobari2025activation,
  title={Activation-informed merging of large language models},
  author={Nobari, Amin Heyrani and Alim, Kaveh and ArjomandBigdeli, Ali and Srivastava, Akash and Ahmed, Faez and Azizan, Navid},
  journal={arXiv preprint arXiv:2502.02421},
  year={2025}
}

@article{liu2025sens,
  title={Sens-merging: Sensitivity-guided parameter balancing for merging large language models},
  author={Liu, Shuqi and Wu, Han and He, Bowei and Han, Xiongwei and Yuan, Mingxuan and Song, Linqi},
  journal={arXiv preprint arXiv:2502.12420},
  year={2025}
}

@article{sukhbaatar2024branch,
  title={Branch-train-mix: Mixing expert llms into a mixture-of-experts llm},
  author={Sukhbaatar, Sainbayar and Golovneva, Olga and Sharma, Vasu and Xu, Hu and Lin, Xi Victoria and Rozi{\`e}re, Baptiste and Kahn, Jacob and Li, Daniel and Yih, Wen-tau and Weston, Jason and others},
  journal={arXiv preprint arXiv:2403.07816},
  year={2024}
}

@article{liu20251bit,
  title={1bit-Merging: Dynamic Quantized Merging for Large Language Models},
  author={Liu, Shuqi and Yao, Yuxuan and He, Bowei and Liu, Zehua and Han, Xiongwei and Yuan, Mingxuan and Wu, Han and Song, Linqi},
  journal={arXiv preprint arXiv:2502.10743},
  year={2025}
}

@article{hines2004modeldb,
  title={ModelDB: a database to support computational neuroscience},
  author={Hines, Michael L and Morse, Thomas and Migliore, Michele and Carnevale, Nicholas T and Shepherd, Gordon M},
  journal={Journal of computational neuroscience},
  volume={17},
  number={1},
  pages={7--11},
  year={2004},
  publisher={Springer}
}

@inproceedings{vartak2016modeldb,
  title={ModelDB: a system for machine learning model management},
  author={Vartak, Manasi and Subramanyam, Harihar and Lee, Wei-En and Viswanathan, Srinidhi and Husnoo, Saadiyah and Madden, Samuel and Zaharia, Matei},
  booktitle={Proceedings of the Workshop on Human-In-the-Loop Data Analytics},
  pages={1--3},
  year={2016}
}

@article{zaharia2018accelerating,
  title={Accelerating the machine learning lifecycle with MLflow.},
  author={Zaharia, Matei and Chen, Andrew and Davidson, Aaron and Ghodsi, Ali and Hong, Sue Ann and Konwinski, Andy and Murching, Siddharth and Nykodym, Tomas and Ogilvie, Paul and Parkhe, Mani and others},
  journal={IEEE Data Eng. Bull.},
  volume={41},
  number={4},
  pages={39--45},
  year={2018}
}

@inproceedings{barreto2024dvc,
  title={DVC in Open Source ML-development: The Action and the Reaction},
  author={Barreto Simedo Pacheco, Lorena and Rahman, Musfiqur and Rabbi, Fazle and Fathollahzadeh, Pouya and Abdellatif, Ahmad and Shihab, Emad and Chen, Tse-Hsun and Yang, Jinqiu and Zou, Ying},
  booktitle={Proceedings of the IEEE/ACM 3rd International Conference on AI Engineering-Software Engineering for AI},
  pages={75--80},
  year={2024}
}

@inproceedings{buneman2001and,
  title={Why and where: A characterization of data provenance},
  author={Buneman, Peter and Khanna, Sanjeev and Wang-Chiew, Tan},
  booktitle={International conference on database theory},
  pages={316--330},
  year={2001},
  organization={Springer}
}

@inproceedings{green2007provenance,
  title={Provenance semirings},
  author={Green, Todd J and Karvounarakis, Grigoris and Tannen, Val},
  booktitle={Proceedings of the twenty-sixth ACM SIGMOD-SIGACT-SIGART symposium on Principles of database systems},
  pages={31--40},
  year={2007}
}

@article{eggers2024automating,
  title={Automating Data Lineage and Pipeline Extraction},
  author={Eggers, Sebastian},
  journal={Proceedings of the VLDB Endowment. ISSN},
  volume={2150},
  pages={8097},
  year={2024}
}

@article{bux2015saasfee,
  title={SAASFEE: scalable scientific workflow execution engine},
  author={Bux, Marc and Brandt, J{\"o}rgen and Lipka, Carsten and Hakimzadeh, Kamal and Dowling, Jim and Leser, Ulf},
  journal={Proceedings of the VLDB Endowment},
  volume={8},
  number={12},
  pages={1892--1895},
  year={2015},
  publisher={VLDB Endowment}
}

@inproceedings{van2017versioning,
  title={Versioning for end-to-end machine learning pipelines},
  author={Van Der Weide, Tom and Papadopoulos, Dimitris and Smirnov, Oleg and Zielinski, Michal and Van Kasteren, Tim},
  booktitle={Proceedings of the 1st Workshop on Data Management for End-to-End Machine Learning},
  pages={1--9},
  year={2017}
}

@article{vadde2024devops,
  title={DevOps in the Age of Machine Learning: Bridging the Gap Between Development and Data Science},
  author={Vadde, Bharath Chandra and Munagandla, VB},
  journal={International Journal of Machine Learning Research in Cybersecurity and Artificial Intelligence},
  volume={15},
  number={1},
  pages={530--544},
  year={2024}
}

@article{ruan2021lineagechain,
  title={LineageChain: a fine-grained, secure and efficient data provenance system for blockchains},
  author={Ruan, Pingcheng and Dinh, Tien Tuan Anh and Lin, Qian and Zhang, Meihui and Chen, Gang and Ooi, Beng Chin},
  journal={The VLDB Journal},
  volume={30},
  number={1},
  pages={3--24},
  year={2021},
  publisher={Springer}
}

@article{chen2021evaluating,
  title={Evaluating large language models trained on code},
  author={Chen, Mark},
  journal={arXiv preprint arXiv:2107.03374},
  year={2021}
}

@article{zhou2023instruction,
  title={Instruction-following evaluation for large language models},
  author={Zhou, Jeffrey and Lu, Tianjian and Mishra, Swaroop and Brahma, Siddhartha and Basu, Sujoy and Luan, Yi and Zhou, Denny and Hou, Le},
  journal={arXiv preprint arXiv:2311.07911},
  year={2023}
}

@inproceedings{dua2019drop,
  title={DROP: A Reading Comprehension Benchmark Requiring Discrete Reasoning Over Paragraphs},
  author={Dua, Dheeru and Wang, Yizhong and Dasigi, Pradeep and Stanovsky, Gabriel and Singh, Sameer and Gardner, Matt},
  booktitle={Proceedings of the 2019 Conference of the North American Chapter of the Association for Computational Linguistics: Human Language Technologies, Volume 1 (Long and Short Papers)},
  pages={2368--2378},
  year={2019}
}

@String{Computing = "Computing" }

@String{Springer = "Springer-Verlag" }

@ArtifactSoftware{R,
    title = {R: A Language and Environment for Statistical Computing},
    author = {{R Core Team}},
    organization = {R Foundation for Statistical Computing},
    address = {Vienna, Austria},
    year = {2019},
    url = {https://www.R-project.org/},
}

\end{document}